\RequirePackage{fix-cm}
\documentclass[smallextended]{svjour3} 
\smartqed 
\usepackage{array}
\usepackage{amssymb}
\usepackage{amsfonts}
\usepackage{amstext}
\usepackage{graphicx}
\usepackage{amscd}
\usepackage{amsmath}
\usepackage{enumitem}
\usepackage{blindtext}
\AtBeginDocument{%
  \setlength{\oddsidemargin}{\dimexpr(\paperwidth-\textwidth)/2-1in}%
  \setlength{\evensidemargin}{\oddsidemargin}%
  \setlength{\topmargin}{%
    \dimexpr(\paperheight-\textheight)/2-\headheight-\headsep-1in}%
}

\newcommand{\tr}{\mathrm{tr}}

\begin{document}

\title{Analyses of the transmission of the disorder from a disturbed environment to a spin chain}

\author{Lucile Aubourg \and David Viennot}
\institute{Institut UTINAM (CNRS UMR 6213, Universit\'e de Bourgogne-Franche-Comt\'e, Observatoire de Besan\c con), 41bis Avenue de l'Observatoire, BP1615, 25010 Besan\c con cedex, France.}

\maketitle

\begin{abstract}
We study spin chains submitted to disturbed kick trains described by classical dynamical processes. The spin chains are described by Heisenberg and Ising models. We consider decoherence, entanglement and relaxation processes induced by the kick irregularity in the multipartite system (the spin chain). We show that the different couplings transmit the disorder along the chain differently and also to each spin density matrix with different efficiencies. In order to analyze and to interpret the observed effects we use a semi-classical analysis across the Husimi distribution. It consists to consider the classical spin orientation movements. A possibility of conserving the order into the spin chain is finally analyzed.
\end{abstract}
                                                                                                                                                                                                                                                                                                                                                                                                                                                                                                                                                                                                                                                                                                                                                                                                                                                                                                                                                                                                                                                                                                                                                                                                                                                                                                                                                                              \section{Introduction}
The quantum dynamics and the quantum control of multipartite quantum systems have attracted much attention due to their applications to quantum information protocols (to perform logic gates and for the transport and the teleportation of information) and to nanosciences (control of small nanostructures). A key problem is the understanding of the dynamical processes associated with the whole multipartite quantum system. These processes have a consequence on each component of the system, and could induce decoherence and relaxation. (\cite{breuer,lages,gedik,lages2,rossini,zhou,castanino,xu,brox}). In order to explore this problem we consider the interesting example of a multipartite quantum system represented by a spin chain, i.e. a set of $N$ $\frac{1}{2}$-spins two by two coupled in order to form an open line chain. In this paper we consider both the Heisenberg and the Ising model to describe the spin-spin interactions which are responsible for the ``cohesion'' of the chain. The dynamical processes on the spin chain are induced by ultra-short kicks on all spins. The subject concerning decoherence processes of regularly kicked spin chains has been studied by some authors \cite{prosen,prosen2,prosen3,prosen4,lakshmin,boness,pineda}. In these previous studies, the kick processes are regular. In the present paper, we want to consider irregular trains of kicks (the strengths and the delays of the different kicks are modified with respect to the time) which can potentially induce richer dynamical behaviors of the spin chain. Indeed in a previous paper \cite{viennot2013}, we have observed several interesting behaviors. In this previous study we have considered only irregular kicked spin ensembles without any coupling between the spins. We cannot then consider this ensemble as a multipartite system (no information is exchanged between the spins) but only as a set of independent systems dephased during the evolution. A goal of this paper is the study of such a phenomenon for a spin chain where the coupling induces inner decoherence and entanglement processes.

\begin{figure}
\begin{center}
\includegraphics[width=7.7cm]{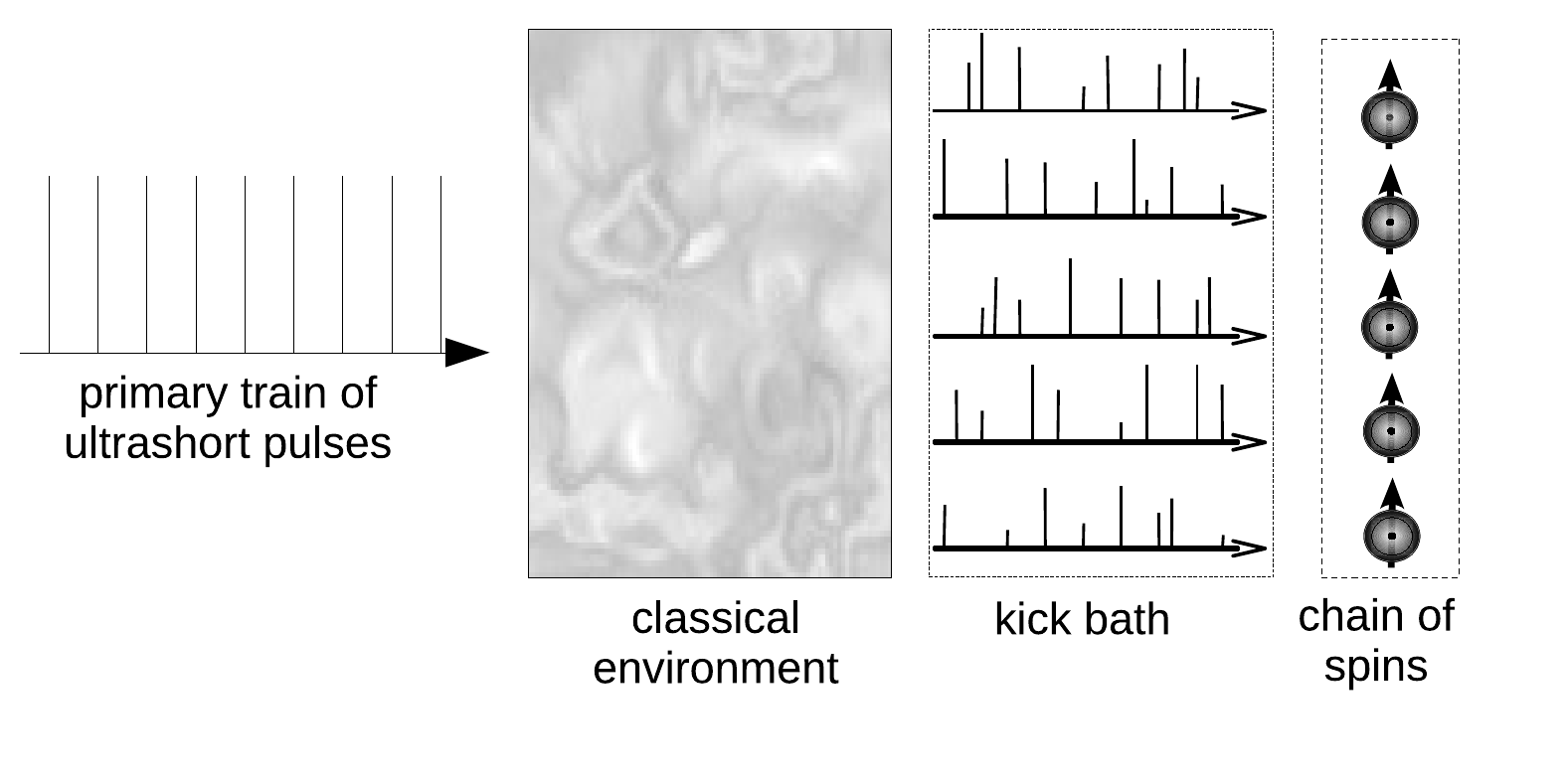}
\caption{\label{kickedspinbath2} Schematic representation of a quantum spin chain controlled by a disturbed train of ultrashort pulses. The set of kick trains issued from the disturbance constitutes a kind of ``classical kick bath''.}
\end{center}
\end{figure}

The motivation to study of irregularly kicked spin chains is that in some situations the primary train of kicks addressed to the spins, must go through an environment (considered as classical in this paper) before to reach the target spins. It can be disturbed by this environment, see fig. \ref{kickedspinbath2}. The disturbance can attenuate the kick strengths and/or delay the arrival kicks. Since  each of the kick trains can be irregular, the spins can feel different trains. The set of kick trains is called a kick bath since we can assimilate the model to a spin chain in contact with a kind of classical bath. The disturbances are described by different classical dynamical systems. In this paper, we consider some academic classical dynamics in order to try to understand the behavior of a spin chain submitted to a real environment in next analyses.

A goal of this paper is to interpret the origins of the different evolutions of a kicked spin chain coupled by Heisenberg or Ising models (decoherence, relaxation and entanglement), but also to see the transmission of the disorder in a kicked spin chain. We see in \cite{viennot2013} that the disorder induces into the classical kick bath by the disturbance is transmitted to the spin chain by the kicks. In this paper we will see that this effect is also present when the spins into the chain are coupled. Besides, the interaction parameter is the source of the disorder transmission along the spin chain. It allows a larger disorder into the spin chain. If there is some disorder which appears into the spin chain, which is in opposition with the interaction process, the spins begin to get entangled with their neighbors. This phenomenon conducts to a lost of the initial information of each spins. In contrast with these analyses, it is important to ask the question of the possibility of keeping some coherence, some order into the spin chain in spite of the kick disorder. All this analysis can be very difficult because of the interactions between the spins, the entanglement and the state superpositions. We have chosen to consider an analysis of the phenomenon based on a semi-classical model of the spins \cite{semiclassical}. This analysis is based on the spin Husimi distribution \cite{husimi} which is the quasiprobability distribution of a quantum state onto the classical phase space. Even if our analysis is based on a semi-classical model for the interpretations, all the simulations are performed with a quantum model. 

This paper is organized as follows. Section 2 presents the spin chain models disturbed by the kicks used in this study. The next section is devoted to the general behavior of the kicked spin chain. We study the entropy, the coherence and the population in order to understand how the disorder is transmitted and its consequences on the spin chain with respect to the coupling. We will see that the Ising-X coupling is the interaction which conduces to the most lost of initial information with a large decoherence. Whereas the Heisenberg interaction can conserve some coherence. Section 4 talks about a possibility of no disorder transmission. In these sections, a chain of only ten spins is considered. But the results do not depend on the number of spins into the chain (the justification is given in appendix A).

\section{Dynamics of kicked spin chains}
We consider a chain of $N$ coupled spins by nearest-neighbor interactions. A constant and uniform magnetic field $\vec{B}$ is applied on the spin chain inducing an energy level splitting by Zeeman effect. We denote by $\frac{\hbar w_1}{2}$ the energy splitting ($w_1$ is the quantum frequency which is identical for each spin). At the initial time $t=0$ the chain is generally coherent (a few incoherent cases are also studied), i.e. all spins are in the same quantum state $|\psi_0\rangle = \alpha |\uparrow \rangle+ \beta |\downarrow \rangle$ ($|\alpha|^2+|\beta|^2=1$ with $\alpha,\beta\not=0$ -- $|\psi_0\rangle$ is a ``Schr\"odinger's cat state'' -- ). For $t>0$ the chain is submitted to a train of ultrashort pulses kicking the spins. We suppose that a classical environment disturbs the pulses such that each spin ``views'' a different train (fig. \ref{kickedspinbath2}).
We denote by $w_0 = \frac{2\pi}{T}$ the kick frequency of the primary train. We suppose that the classical environment can attenuate kick strengths and can delay kicks. We denote by $\lambda_n^{(i)}$ and by $\tau_n^{(i)}$ the strength and the delay of the $i$-th kick on the $n$-th spin of the chain. Let ${H_{0_n} = \mathrm{id}^{\otimes (n -1)} \otimes \frac{\hbar \omega_1}{2} |\downarrow \rangle \langle \downarrow|} \otimes \mathrm{id}^{\otimes (N-n)}$ be the quantum Hamiltonian of the $n$-th spin with the Zeeman effect (where we have removed a constant value without significance) and $H_I$ be the nearest-neighbor interaction Hamiltonian which can be for the $n$-th spin of the chain one of the following operators :
\begin{enumerate}
\item Heisenberg coupling 
\begin{equation}
\label{heisenberg}
H_{I_n} =  -J \mathrm{id}^{\otimes (n -1)} \otimes (S_x \otimes S_x + S_y \otimes S_y + S_z \otimes S_z) \otimes \mathrm{id}^{\otimes (N-n-1)}
\end{equation}
with $S_i= \frac{\hbar}{2}\sigma_i$, $\{\sigma_{i}\}_{i=x,y,z}$ are the Pauli matrices and $\mathrm{id^{\otimes n}}$ is the tensor product of ``$n$'' identity matrices of order two.
\item Ising-Z coupling 
\begin{equation}
\label{isingz}
H_{I_n} = -J \mathrm{id}^{\otimes (n-1)} \otimes S_z \otimes S_z
\otimes \mathrm{id}^{\otimes (N-n-1)}
\end{equation}
\item Ising-X coupling 
\begin{equation}
\label{isingx}
H_{I_n} = -J \mathrm{id}^{\otimes (n-1)} \otimes S_x \otimes S_x \otimes \mathrm{id}^{\otimes (N-n-1)}
\end{equation}
\end{enumerate} 
The quantum Hamiltonian of the kicked spin chain is
\begin{equation}
H(t) = \sum_{n=1}^N \Big( H_{0_n} + H_{I_n} + \mathrm{id}^{\otimes (n-1)}
\otimes \hbar W \sum_{i \in \mathbb N} \lambda_n^{(i)} \delta\left(t- iT+\tau_n^{(i)} \right)\otimes \mathrm{id}^{\otimes (N-n)} \Big)
\end{equation}
where $\delta(t)$ is the Dirac distribution and where the kick operator $W$ is a rank one projector: $W = |w \rangle \langle w|$ with the kick direction ${|w \rangle = \cos \vartheta |\uparrow \rangle + \sin \vartheta |\downarrow \rangle}$ (for the sake of simplicity we do not consider a relative phase between the two components of $|w \rangle$). By considering the reduced time $\theta = \frac{2\pi t}{T} = w_0 t$ we have
\begin{equation}
\label{dynamics}
H(\theta) = \sum_{n=1}^N \Big( H_{0_n} + H_{I_n} + \mathrm{id}^{\otimes (n -1)} \otimes \hbar W \sum_{i \in \mathbb N} \lambda_n^{(i)} \delta\left(\theta - 2i \pi +\varphi_n^{(i)} \right)\otimes \mathrm{id}^{\otimes (N-n)} \Big)
\end{equation}
with the angular delay $\varphi_n^{(i)} = w_0 \tau_n^{(i)}$. The $i$-th monodromy operator (the evolution operator from $t=\frac{2 i \pi}{w_0}$ to $\frac{2(i+1)\pi}{w_0}$) is, if the spins are organized from the smallest delay (for $n=1$) to the greatest one (for $n=N$) \cite{viennot}:
\begin{multline}
\label{monodromy}
U^{(i)} = e^{-\frac{\imath H_{0,I}}{\hbar w_0} (2\pi - \varphi_N^{(i)})} \prod_{n=1}^N \left[ \mathrm{id}^{\otimes (N-n)}\right. \otimes (id+(e^{-\imath \lambda_{N-n+1}^{(i)}}-1)W)\\
\otimes \mathrm{id}^{\otimes (n-1)} \left. \times e^{-\frac{\imath H_{0,I}}{\hbar \omega_0} (\varphi_{N-n+1}^{(i)} - \varphi_{N-n}^{(i)})}\right] e^{-\frac{\imath H_{0,I}}{\hbar w_0} \varphi_1^{(i)}}
\end{multline}
with $H_{0,I} = \sum_{n=1}^N (H_{0_n} + H_{I_n})$.
We see that the monodromy operator is $2\pi$-periodic with respect to the kick strength; $\lambda_n^{(i)}$ is then defined modulo $2\pi$ from the viewpoint of the quantum system. Thus the strength-delay pair $(\lambda,\varphi)$ defines a point on a torus $\mathbb T^2$ which plays the role of a classical phase space for the kick train. The classical dynamics used for the classical environment after the first kicks are
\begin{enumerate}
\item Stationary bath defined by the flow
\begin{equation}
\Phi \begin{pmatrix}
\lambda \\
\phi
\end{pmatrix} = \begin{pmatrix}
\lambda \\
\phi
\end{pmatrix}
\end{equation}
\item Drifting bath defined by the flow
\begin{equation}
\Phi \begin{pmatrix}
\lambda \\
\phi
\end{pmatrix} = \begin{pmatrix}
\lambda + \frac{2\pi}{a} \ \ \ \ mod(2\pi)\\
\phi + \frac{2\pi}{a} \ \ \ \ mod(2\pi)
\end{pmatrix}
\end{equation}
where $a,b \in \mathbb{R} \backslash \mathbb{Q}$. The orbit of ($\lambda_0 , \varphi_0 $) by $\Phi$ is dense on $\mathbb{T}^2$.
\item Microcanonical bath defined by a flow consisting to random variables on $\mathbb{T}^2$, with the uniform probability measure
\begin{equation}
d\mu(\lambda, \varphi) = \frac{d\lambda d\varphi}{4 \pi^2}
\end{equation}
where $\mu$ is the Haar probability measure on $\mathbb{T}^2$.
\item Markovian bath defined by a stochastic flow consisting to random variables on $\mathbb{T}^2$ with the following probability measure
\begin{equation}
d \nu_n(\lambda, \varphi) =  \frac{d\lambda d\varphi}{\sqrt{2 \pi \sigma}} e^{-\frac{1}{2 \sigma}((\lambda - \lambda_{n-1})^2 + (\varphi - \varphi_{n-1})^2)}
\end{equation}
This process is a discrete-time Wiener process (a random walk) corresponding to a Brownian motion on $\mathbb{T}^2$ with average step equal to $\sigma > 0$.
\end{enumerate}

Kick baths are defined also by the initial distribution of the first kicks $\{ (\lambda_n^{(0)},\varphi_n^{(0)}) \}_{n=1,...,N}$. These first kicks are randomly chosen in $[\lambda_*,\lambda_*+d_0] \times [\varphi_*,\varphi_*+d_0]$ (with uniform probabilities). $(\lambda_*,\varphi_*)$ can be viewed as the parameters of the primary kick train. The length of the support of the initial distribution (the initial dispersion) $d_0$ is the magnitude of the disturbance on the first kick.\\

Let $|\psi^{(i)} \rangle \in \mathbb{C}^{2N}$ be the state of the chain at time $t=iT$ ($|\psi^{(i)} \rangle$ represents the ``stroboscopic'' evolution of the chain). By definition of the monodromy operator we have
\begin{equation}
|\psi^{(i+1)} \rangle = U^{(i)} |\psi^{(i)} \rangle
\end{equation}
The density matrix of the chain is then
\begin{equation}
\rho^{(i)} = \frac{1}{N} |\psi^{(i)}\rangle \langle \psi^{(i)}|
\end{equation}
and the density matrix of the n-th spin is
\begin{equation}
\rho_{n}^{(i)} = Tr_{i=1,...,n-1,n+1,...,N}(\rho^{(i)})
\end{equation}
$ Tr_{i=1,...,n-1,n+1,...,N}$ is the partial trace on all the spin Hilbert spaces except the $n$-th one. It encodes two fundamental informations:
\begin{enumerate}[label=\roman*]
\item the populations $\langle \uparrow| \rho_{n}^{(i)} |\uparrow \rangle$ and $\langle \downarrow |\rho_{n}^{(i)}|\downarrow \rangle$ which are the occupation probabilities of the states $|\uparrow \rangle$ and $|\downarrow \rangle$ for the $n$-th spin.
\item the coherence $|\langle \uparrow| \rho_n^{(i)} |\downarrow \rangle|$ which measures the coherence of the $n$-th spin of the chain \cite{breuer,bengtsson}.
\end{enumerate}

We deduce from $\rho_{n}^{(i)}$, the density matrix of the average spin of the chain for the $i$-th kick
\begin{equation}
\rho_{tot}^{(i)} = \frac{1}{N} \sum_{n=1}^N \rho_{n}^{(i)}
\end{equation}

In the next section, we want to know the general behavior of the spin chain coupled by the three interactions (see above eq. \ref{heisenberg}, \ref{isingz} and \ref{isingx}). In addition, the chain is submitted to stationary, Markovian, drifting and microcanonical kick baths for small or large dispersions of the control parameters (strength and delay) $d_0$, on the phase space (the torus). This last parameter quantifies the initial kick dispersion on all spins of the chain. More this parameter is large, more the kicks received by two spins are different. So, it induces an initial disorder into the kick bath. Firstly by the use of the entropy, we will quantify the transmission of the disorder to the spin chain. Secondly, we study the effects of the disorder on the population and on the coherence for one spin and for an average spin (to see if one spin has the same behavior than the whole chain). After we are interested by the understanding of the evolution of a spin chain induced by the kicks and the various couplings, using the previous analyses and the Husimi distribution. Finally, in another section, we analyze a possibility to not induced a transmission of the disorder to the spin chain.

\section{General evolution of a kicked spin chain} 
In order to understand the evolution of the kicked spin chain, we introduce the notions of entropy, relaxation and decoherence.

One of the main physical phenomenon in this system is related to the production and the transmission of the disorder. The dynamics of the kick bath produces disorder which is transmitted to the spin chain. An increase of the chain entropy corresponds to an increase of the disorder into the chain. 
We consider the von Neumann entropy of the spin chain : 
\begin{equation}
S_{vN} (\rho) =-\frac{1}{\ln(2)}tr(\rho \log \rho)
\end{equation}
$tr$ corresponds to the matricial trace, $\log$ denotes the matricial natural logarithm and $\frac{1}{\ln(2)}$ is just a normative factor (which allows a maximal entropy of 1). The von Neumann entropy of one spin $S_{vN} (\rho_n)$ measures the entanglement of the $n$-th spin with the other ones of the chain. But, $S_{vN} (\rho_{tot})$, the von Neumann entropy of the average on the whole chain is a measure of the disorder into the chain in accordance with the interpretations of the statistical mechanics. In the both cases, the von Neumann entropy is a measure of a lack of knowledge about the system. For one spin ($S_{vN}(\rho_n)$) the lack of knowledge is induced by its entanglement with the other spins. For the whole chain ($S_{vN}(\rho_{tot})$) the lack of knowledge is induced by the fact that the state of a spin randomly chosen in the chain is unknown due to the disorder of the chain modelled by the statistical distribution of state average $\frac{1}{N} \frac{1}{\ln(2)} \sum_{n=1}^N \rho_n$.

The decoherence corresponds to a decrease of the spin coherence $|\langle \uparrow |\rho_n ^{(i)}| \downarrow \rangle|$ into the chain with the number of kicks $i$. The decoherence process is complete if $\underset{i \to + \infty}{\lim} |\langle \uparrow |\rho_n ^{(i)}| \downarrow \rangle| = 0 $. The relaxation corresponds with a loss of the memory of the initial state $|\psi_0 \rangle$. For a maximal relaxation, the spin evolution is really close to the microcanonical distribution $\left( \rho = \begin{pmatrix}
\frac{1}{2} & 0 \\
0 & \frac{1}{2}
\end{pmatrix} \right)$. We analyze both the population and coherence evolution of one spin, and the evolution of an average spin (averaged on all populations and coherences of the spins of the chain at each kick).

\subsection{Parameters responsible for the disorder transmission}
\begin{figure}
\begin{center}
\includegraphics[width=7.7cm]{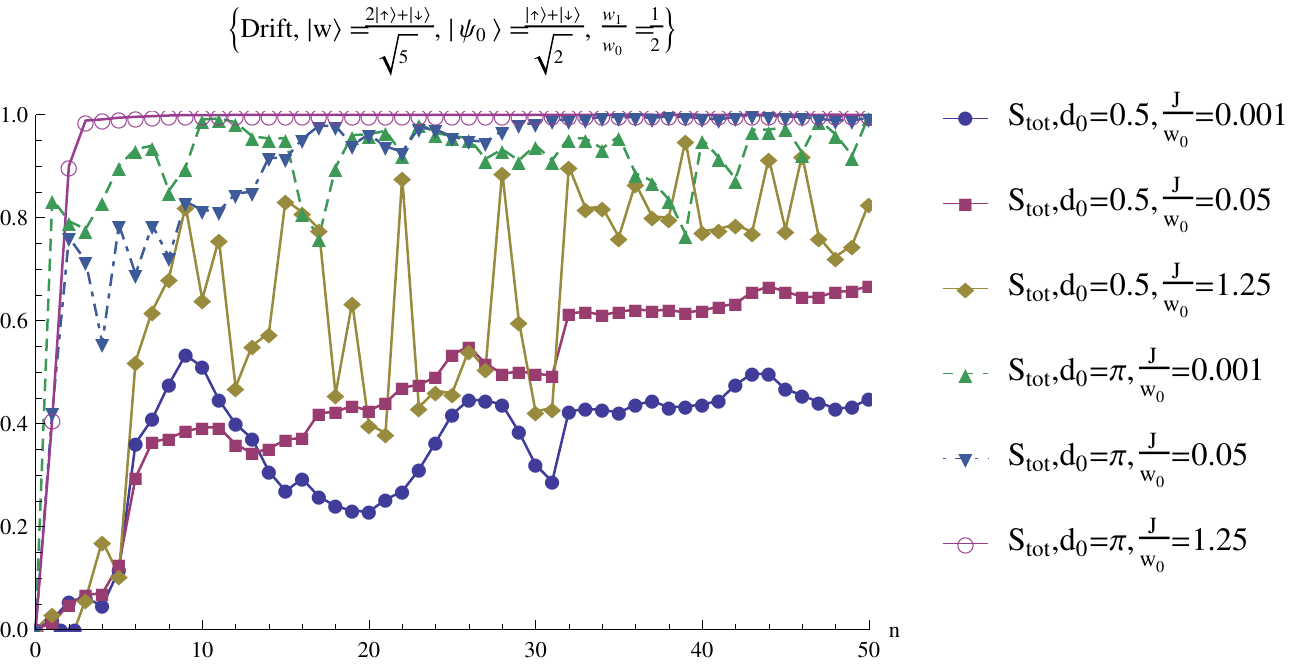}
\includegraphics[width=7.7cm]{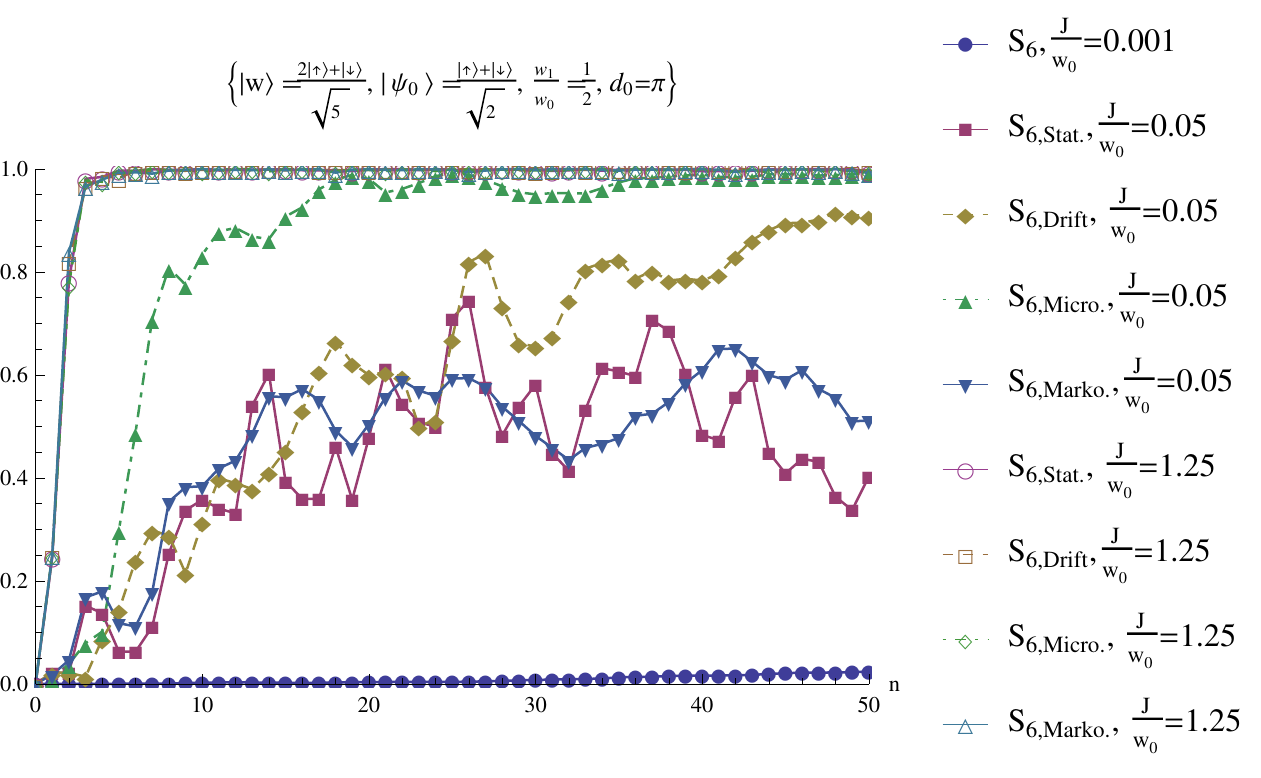}
\caption{\label{driftisingxentropyd0} Evolutions of the entropy of a ten spin chain (up)	and of the entanglement of the sixth spin of the chain with respect to J. Each spin is submitted to a drift (up and down), a stationary, a Markovian and a microcanonical (down) kick bath and is coupled by the Heisenberg, Ising-Z or Ising-X interaction. For the Markovian kick bath, $\sigma= 10^{-3}$.}
\end{center}
\end{figure}
\begin{figure}
\begin{center}
\includegraphics[width=7.7cm]{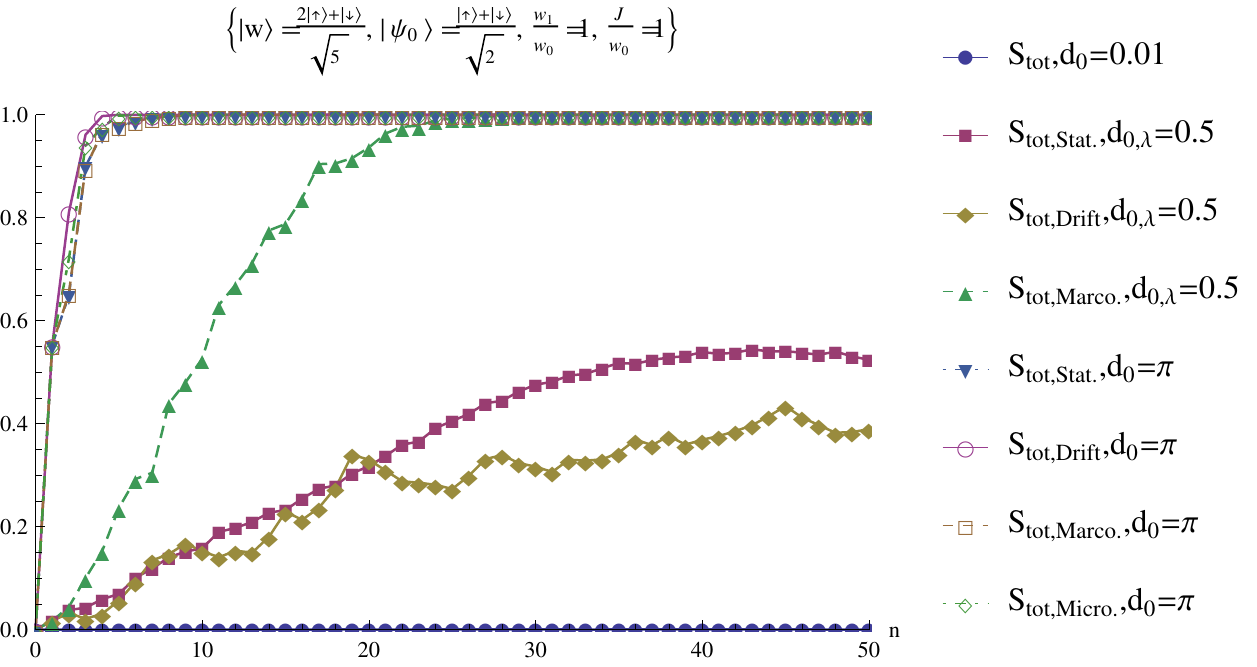}
\includegraphics[width=7.7cm]{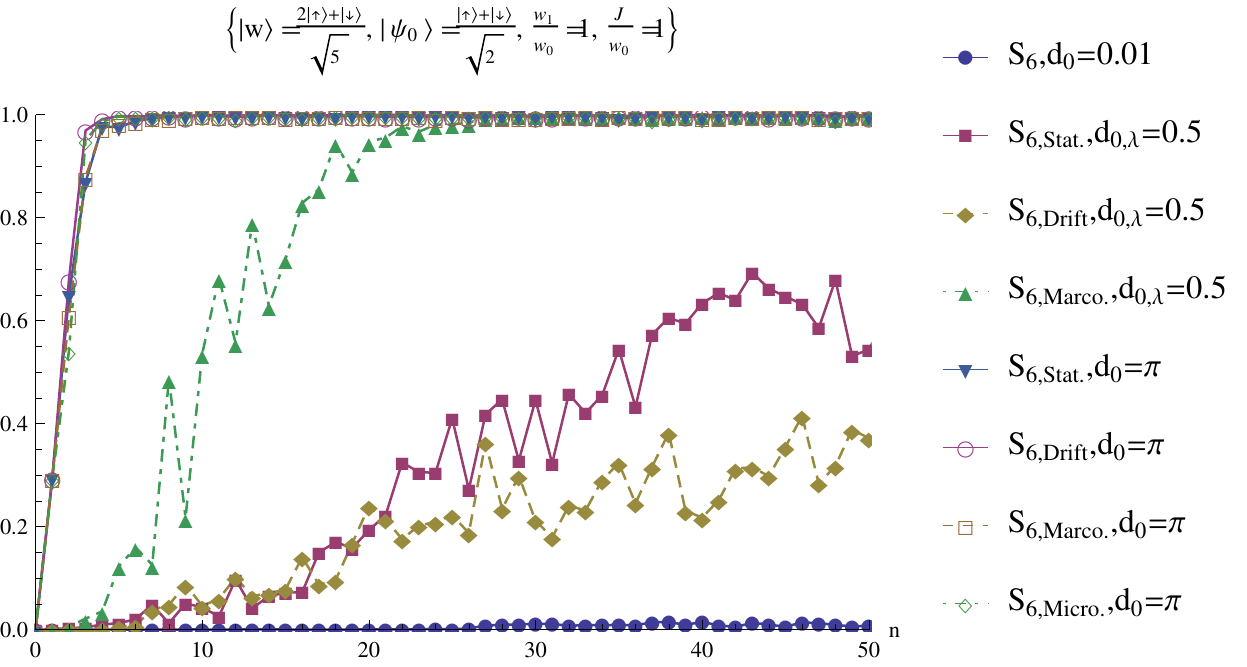}
\caption{\label{entropyheisenbergd0} Evolution of the entropy of a ten spin chain (up) and the entanglement of the fifth spin of the chain (down) with respect to the increase of the initial dispersion. Each spin is submitted to the stationary, the drift, the Markovian or the microcanonical kick bath and is coupled to its neighbors by the Heisenberg interaction. The initial conditions are the same for all dynamics. The Markovian kick bath is characterized by an average Brownian step on the torus of $\sigma=0.1$. The first curve named by $S_{tot},d_0 = 0.01$ is the same for all classical dynamics. $d_{0,\lambda}$ corresponds to an initial dispersion only on the strength parameter ($\lambda$).}
\end{center}
\end{figure}

\begin{figure}
\begin{center}
\includegraphics[width=7.7cm]{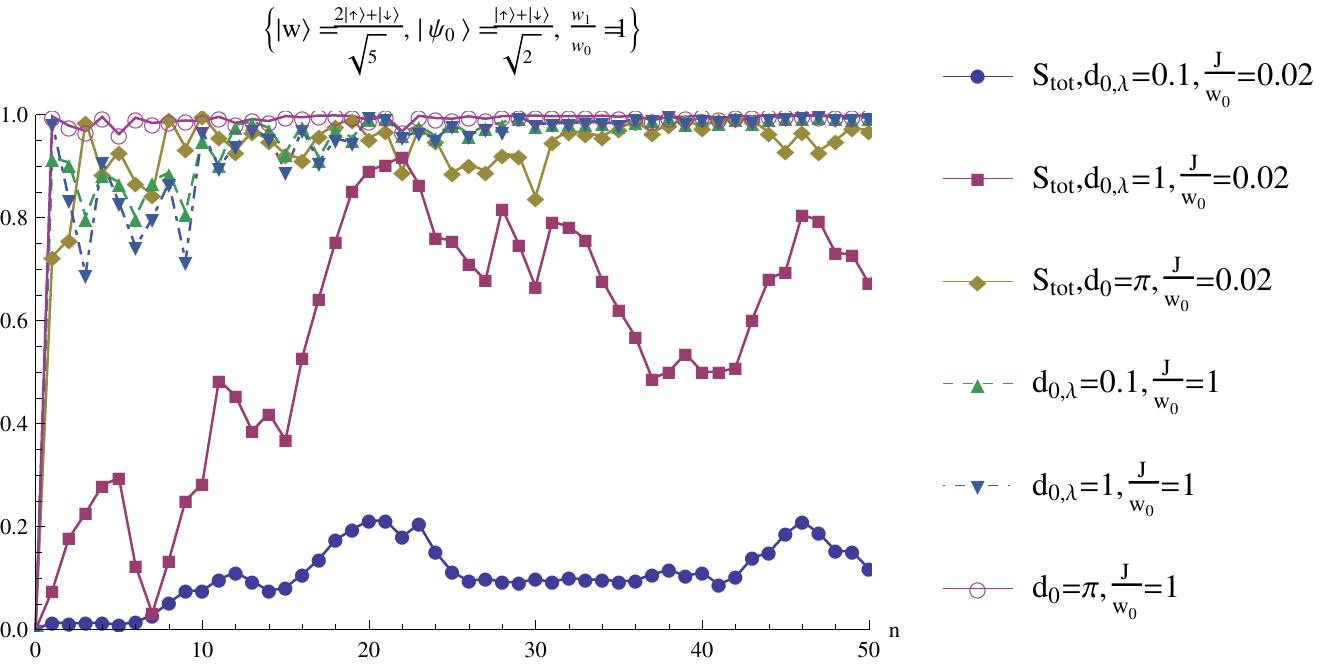}
\includegraphics[width=7.7cm]{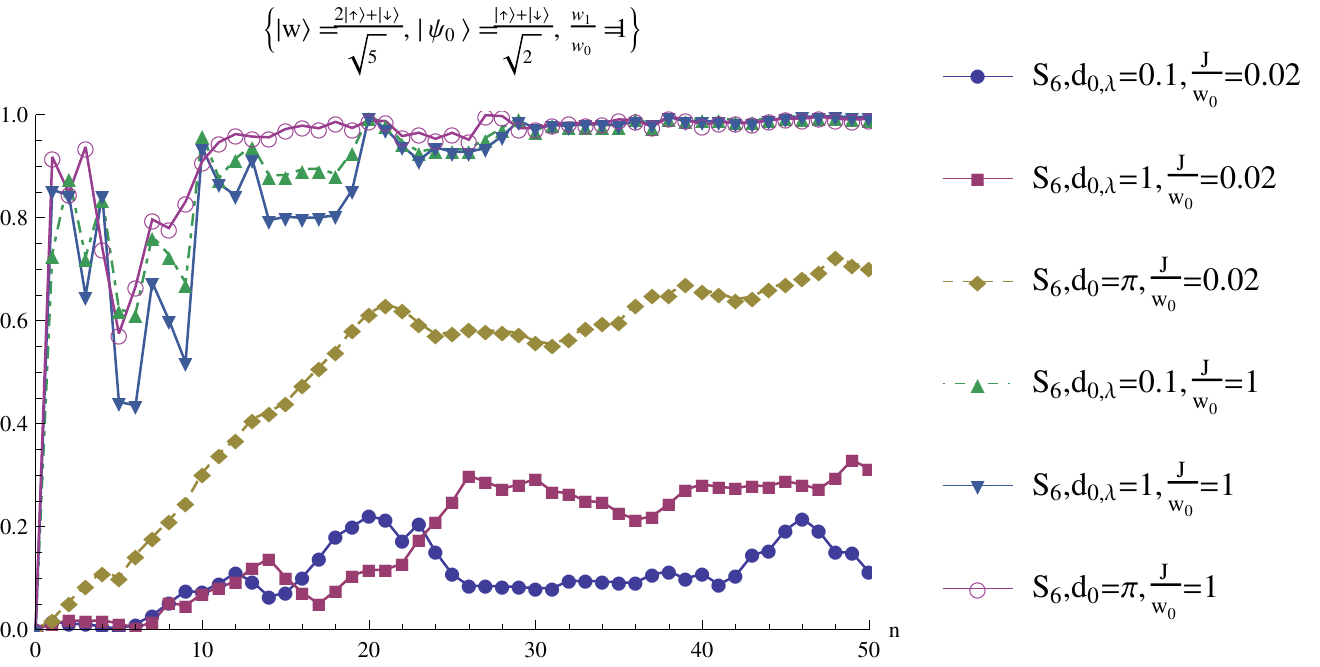}
\caption{\label{isingzentropyavecd0} Evolution of the entropy, up of a ten spin chain ($S_{tot}$) and down of the sixth spin of the chain ($S_6$) with the variation of the kick strengths and of the kick delays ($d_0$) for two values of $\frac{J}{w_0}$. The spins of the chain are submitted to the drift kick baths and are coupled by the Ising-Z interaction. $d_{0,\lambda}$ corresponds to an initial dispersion only on the strength parameter ($\lambda$).}
\end{center}
\end{figure}

\begin{figure}
\begin{center}
\includegraphics[width=7.7cm]{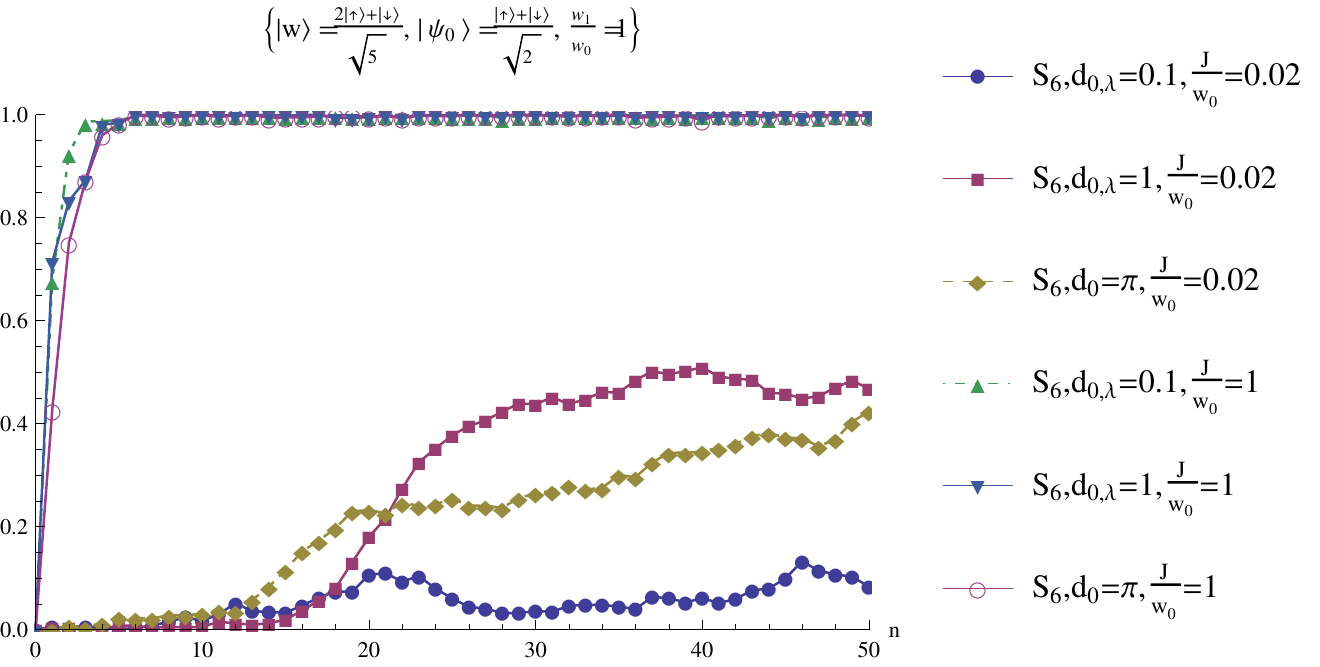}
\caption{\label{driftisingxentropyJ} Evolutions of the entropy of the sixth spin of a ten spin chain coupled by the Ising-X interaction with respect to $d_0$ for two values of $\frac{J}{w_0}$. Each spin is submitted to a drift kick bath. $d_{0,\lambda}$ corresponds to an initial dispersion only on the strength parameter ($\lambda$).}
\end{center}
\end{figure}

The train pulses (the kicks) are disturbed by a classical environment. This last modifies the strength and the delay of each kick. According to the classical environment chosen, the variation of the strength and of the delay between two kicks is different. More a variation between two kicks is large, more the disorder into the kick bath increases. The stationary and the drift bath are less disorder than the microcanonical one. In addition an initial disorder, a variation of the first kick of each train, can be induced by the parameter $d_0$. The initial analysis is to understand how this disorder is transmitted to the spin chain with respect to the interaction. In order to quantify the disorder which appears in the kick chain we use the entropy function. 

We consider a chain of ten spins coupled by a nearest-neighbor Heisenberg, Ising-Z or Ising-X interaction. Figure \ref{driftisingxentropyd0} represents the evolution of the entropy of a ten spin chain (up) and the evolution of the entanglement of one spin of the chain (down). These graphics are the same for all the interactions. The entropy graphic shows that an increase of the value of the interaction parameter induces an increase of the entropy rate into the spin chain. In addition, more the interaction parameter increases, more the entanglement increase rate is large. The interaction parameter is one of the source of the disorder into the spin chain. It is not at the origin of the disorder but allows a better transmission of it.\\

As for the $d_0$ parameter, the analyses are more difficult because the entropy evolution depends on the interaction. Consider firstly the Heisenberg interaction and fig. \ref{entropyheisenbergd0}. In this figure, the up graphic is for the whole chain and the down one represents the entanglement evolution of only one spin of the chain. We see that the disorder into the chain increases approximately with the same rate than the entanglement of one spin with its neighbors. The increase rate of the entropy and of the entanglement increase more rapidly with a large disorder bath. In other words, the drift bath and the classical one have not an increase rate as rapidly as a microcanonical one which is completely disordered.

We consider now a spin chain coupled by the Ising-Z interaction. This coupling produces an entropy and an entanglement which rapidly increase with respect to $d_0$ and $J$ as we can see fig. \ref{isingzentropyavecd0}. This figure shows (up) the evolution of the entropy of a ten spin chain and (down) the evolution of the entanglement of only one spin of the chain. The increase rate of the entropy and of the entanglement is larger than for the Heisenberg interaction. The entropy for an Ising-X interaction evolves similarly than for Ising-Z coupling (the up graphic of fig.\ref{isingzentropyavecd0}). The entanglement is lightly different. The initial oscillations present for $\frac{J}{w_0}=1$ on the down graphic of fig.\ref{isingzentropyavecd0} for an Ising-Z coupling do not exist for an Ising-X coupling (see fig. \ref{driftisingxentropyJ}). \\

Thus, the initial dispersion and the interaction parameter are the sources of the increase of the entropy and of the entanglement into the spin chain. The increase of these both parameters induce an increase of the disorder into the chain. The dispersion parameter and the classical bath chosen induces a kick bath disorder which is transmitted to each spin of the chain. The interaction parameter allows a transmission of the disorder from one spin to its neighbors. 

From these analyses, we want to know the effect of the disorder on the population and on the coherence of the spin chain.

\begin{figure}
\begin{center}
\includegraphics[width=7.7cm]{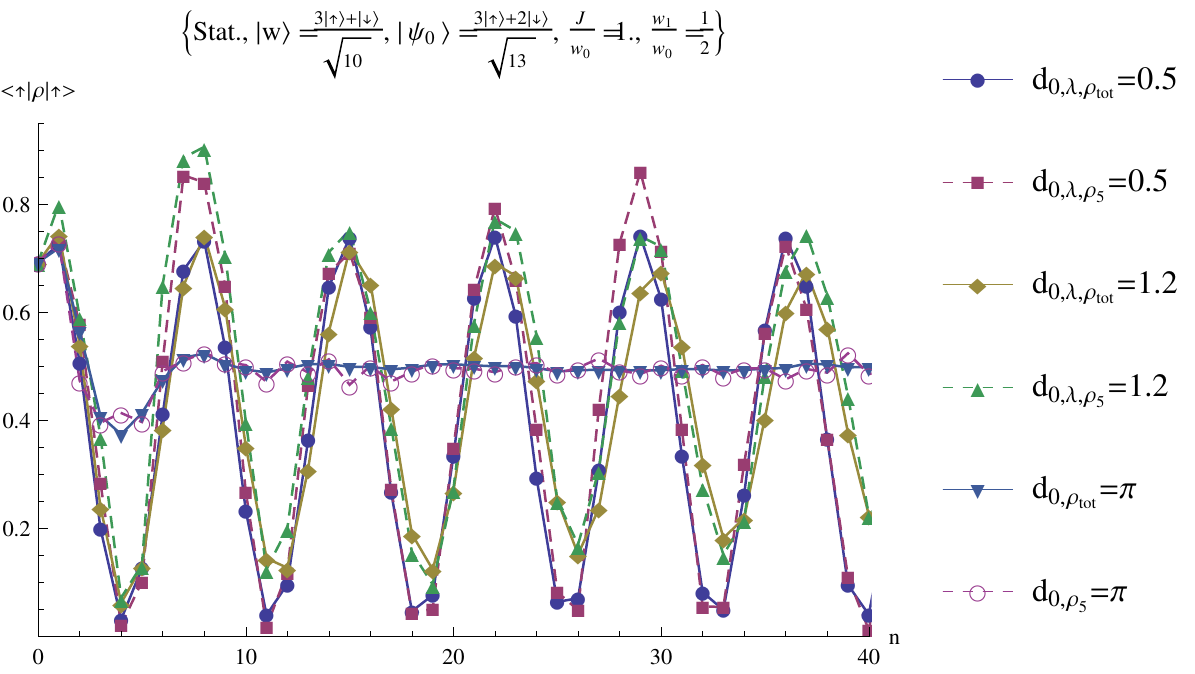}
\includegraphics[width=7.7cm]{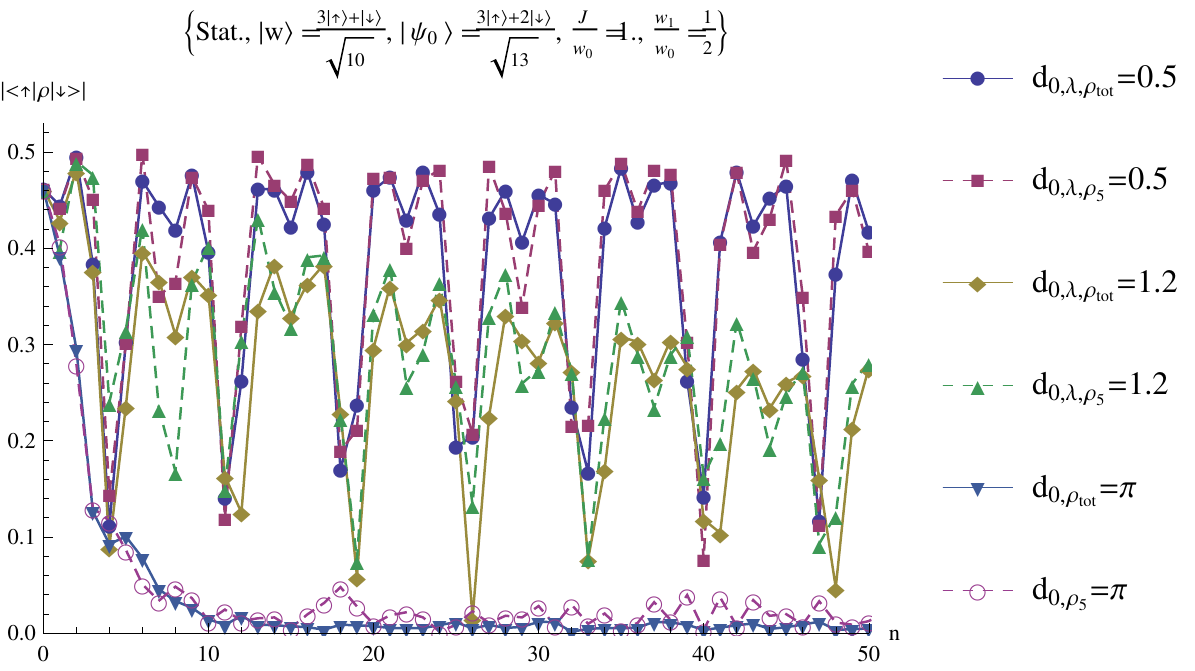}
\caption{\label{popcoheheisenbergaveccouplage} Evolutions of the population (up) and of the coherence (down) of a ten spin chain and of the fifth spin of the chain with an increase of the initial dispersion. Each spin is submitted to a stationary kick bath and the spins of the chain are coupled by a nearest-neighbor Heisenberg interaction. $d_{0,\lambda}$ corresponds to an initial dispersion only on the strength parameter ($\lambda$).}
\end{center}
\end{figure}

\subsection{Population and coherence evolution}

\begin{figure}
\begin{center}
\includegraphics[width=7.7cm]{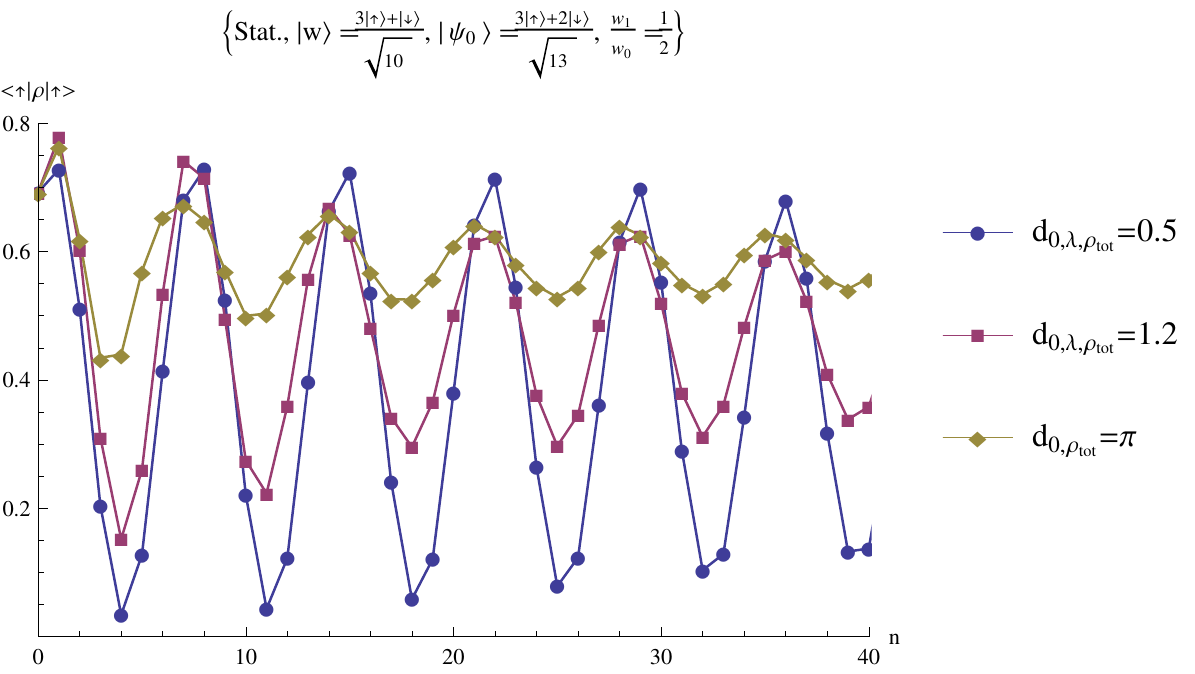}
\includegraphics[width=7.7cm]{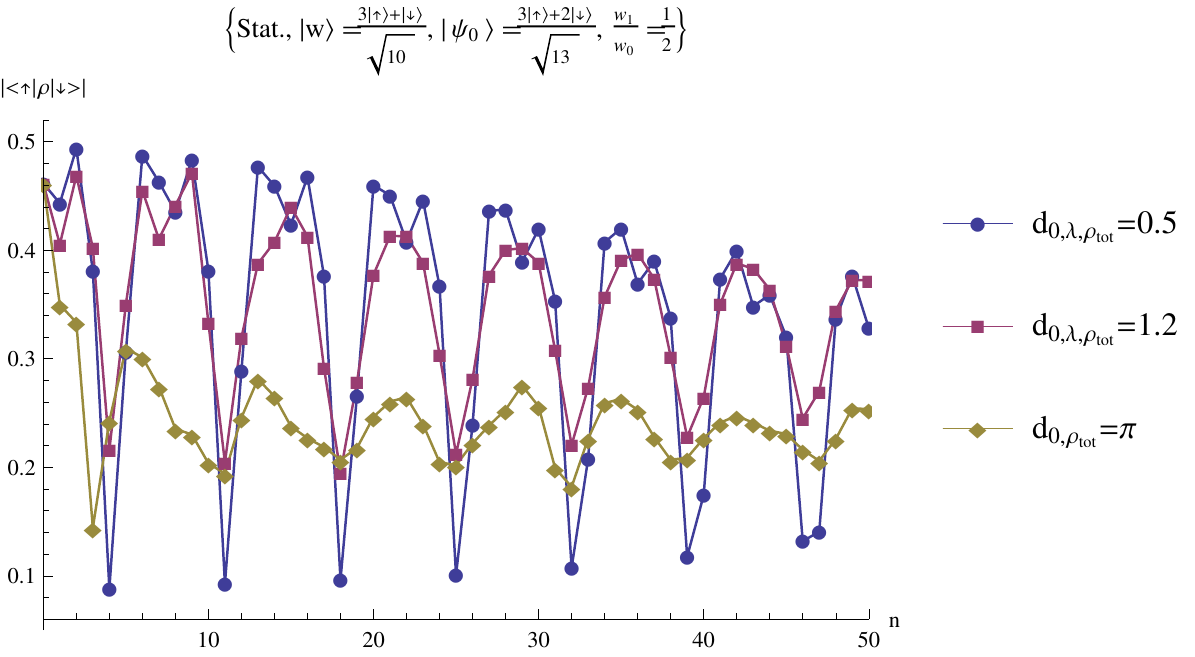}
\includegraphics[width=7.7cm]{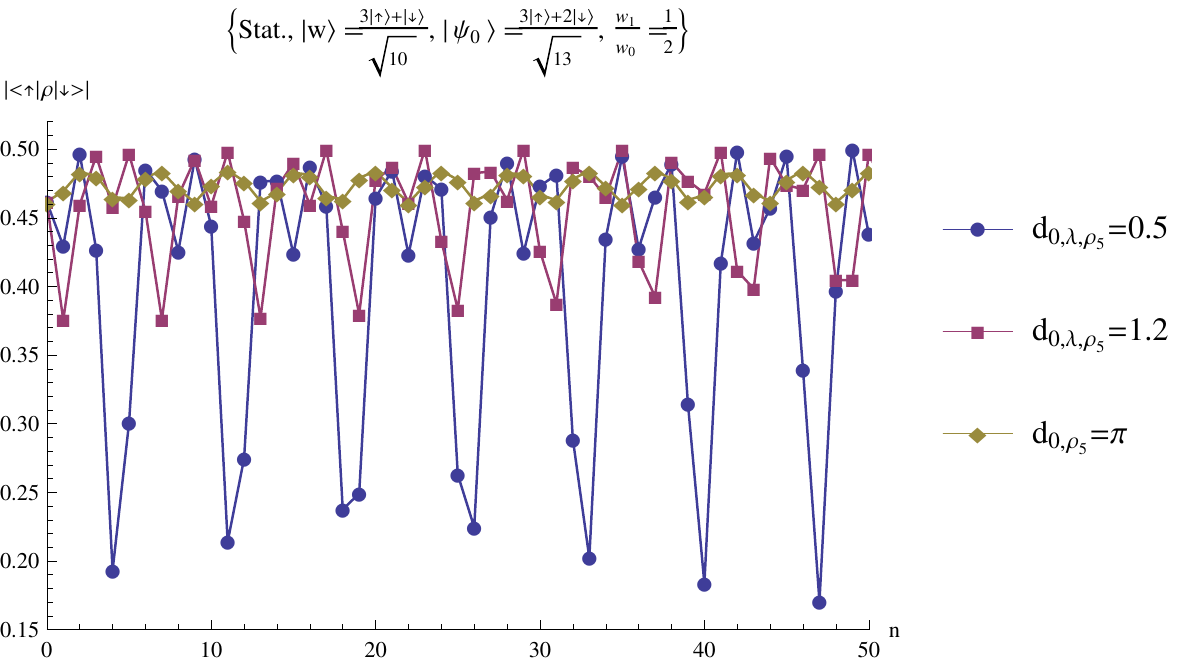}
\caption{\label{popcoheheisenbergsanscouplage} Evolutions of the population and of the coherence of an ensemble of one thousand spins ($ \rho_{tot}$) and of one spin of the ensemble ($\rho_5$) with an increase of the initial dispersion. Each spin is submitted to a stationary kick bath. There is no coupling into the spin chain. $d_{0,\lambda}$ corresponds to an initial dispersion only on the strength parameter ($\lambda$).}
\end{center}
\end{figure}

\begin{figure}
\begin{center}
\includegraphics[width=7.7cm]{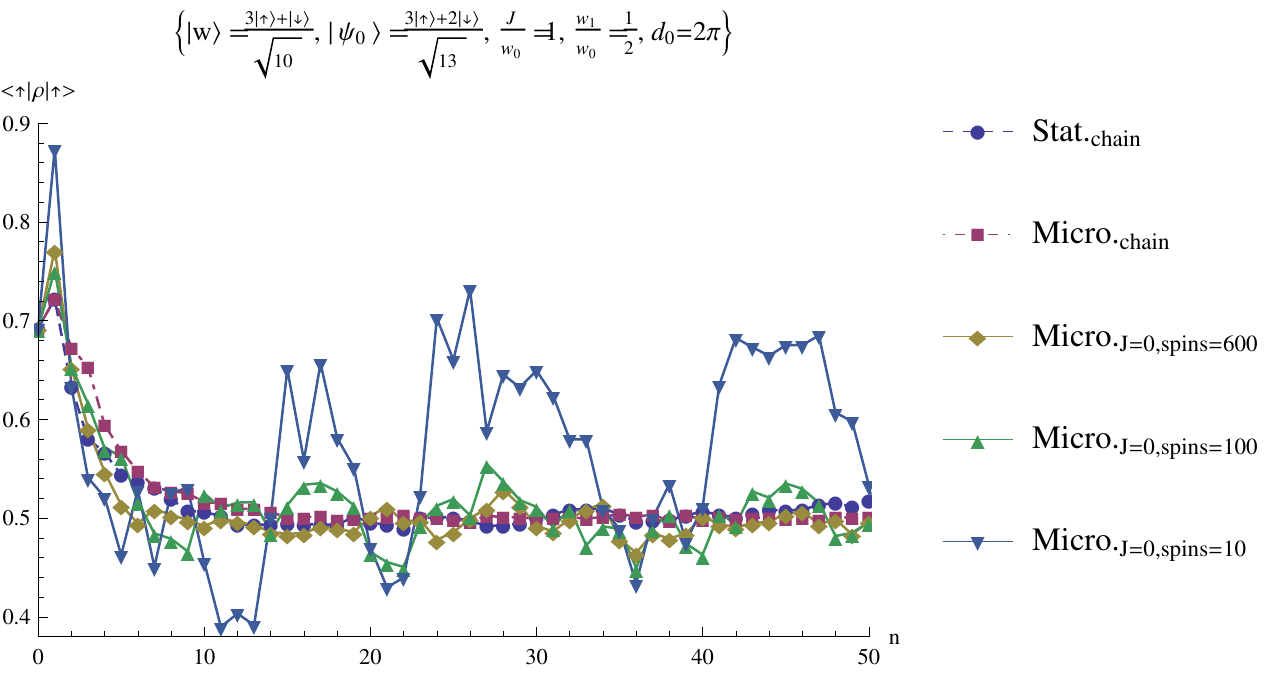}
\includegraphics[width=7.7cm]{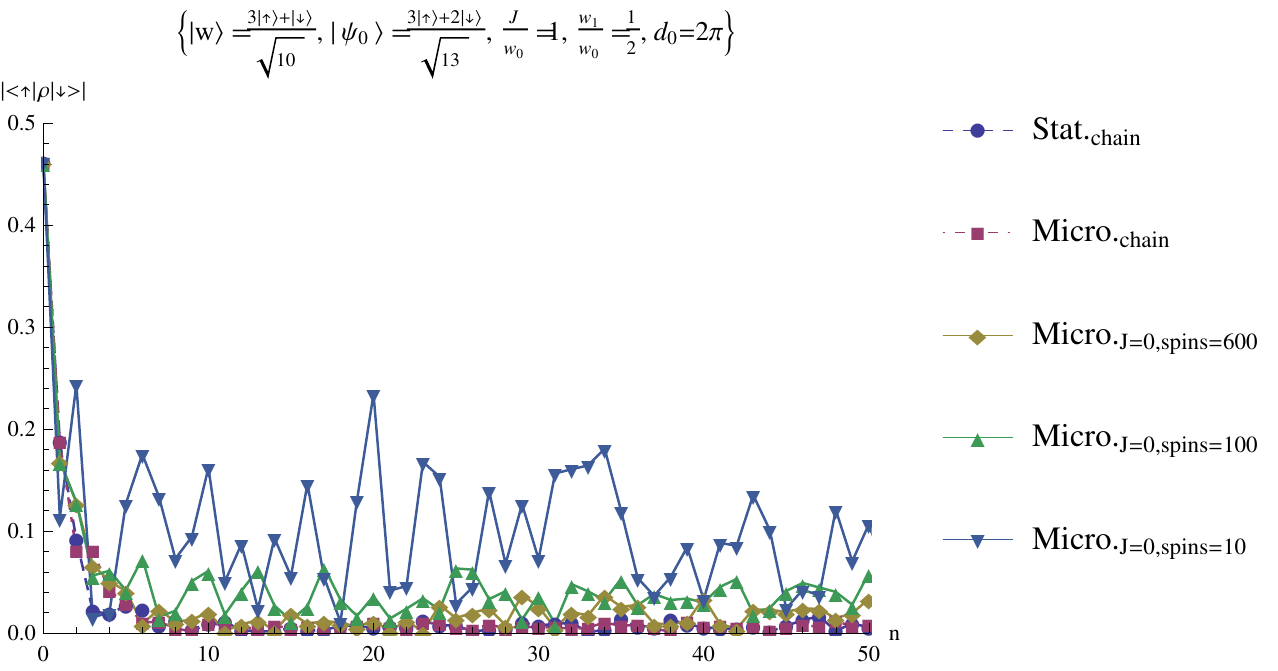}
\caption{\label{popcoheheisenbergavecetsanscouplage} Evolution of the population (up) and of the coherence (down) of a ten spin chain and of a spin ensemble without coupling. The ensemble is kicked by a microcanonical kick bath and the chain by a microcanonical or a stationary one. The spins of the chain are coupled by a nearest-neighbor Heisenberg interaction.}
\end{center}
\end{figure}

\begin{figure}
\begin{center}
\includegraphics[width=7.7cm]{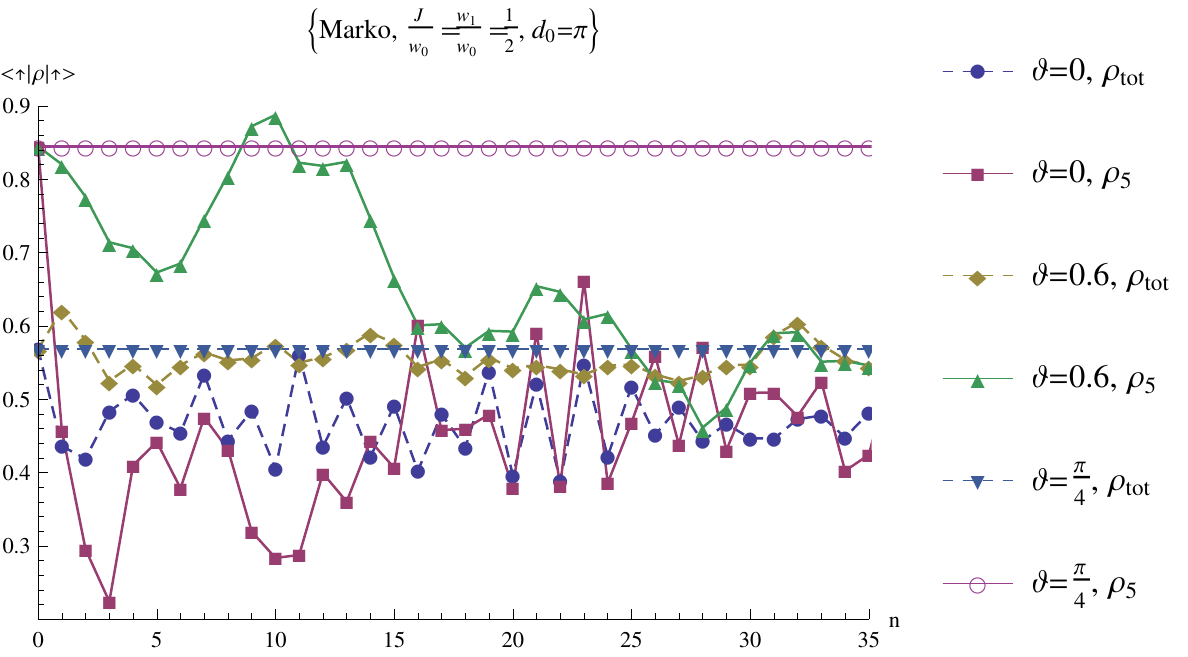}
\caption{\label{isingzvecteurpropre} Evolution of the population of a ten spin chain ($\rho_{tot}$) and of the fifth spin of the chain ($\rho_5$). Each spin is submitted to a Markovian kick bath (with $\sigma = 0.1$) and is coupled by the Ising-Z interaction. The initial state of each spin is randomly chosen to obtain $|\psi_0 > = a |\uparrow > + b |\downarrow >$, with $a \in [0.4,1]$ and $\sqrt{a^2 + b^2}=1$. The kick direction is characterized by $|w>=\cos(\frac{\pi}{4} - \vartheta)|\uparrow> + \sin(\frac{\pi}{4} - \vartheta) |\downarrow>$.}
\end{center}
\end{figure}

\begin{figure}
\begin{center}
\includegraphics[width=7.7cm]{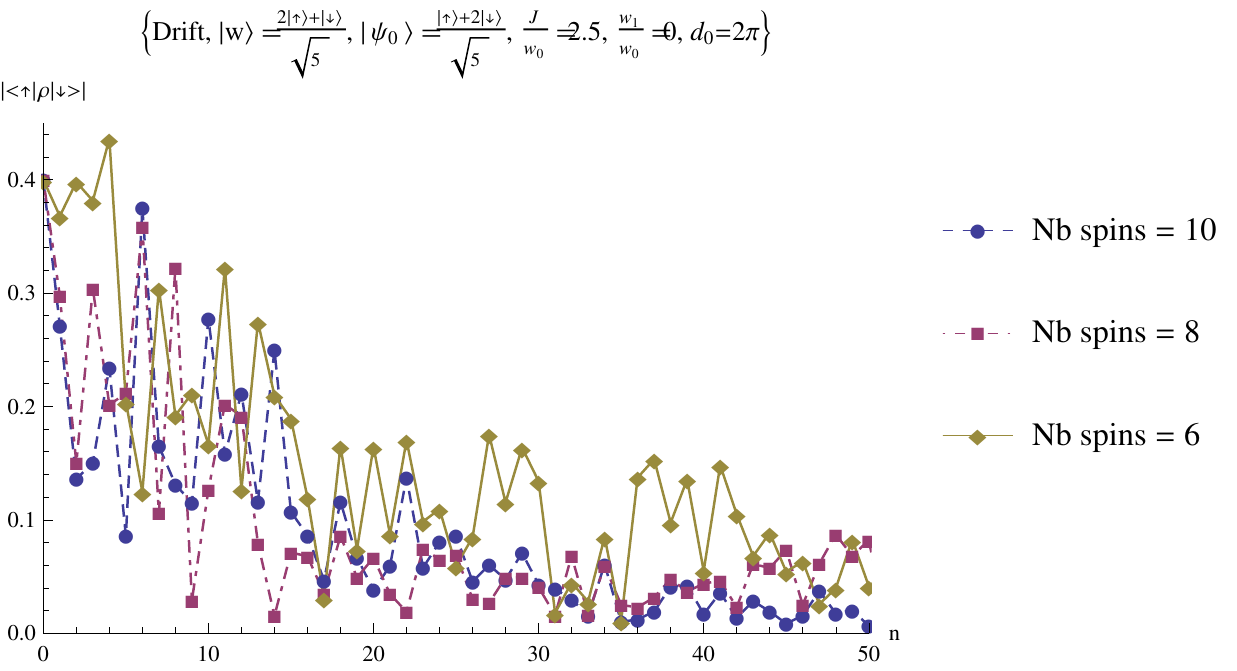}
\includegraphics[width=7.7cm]{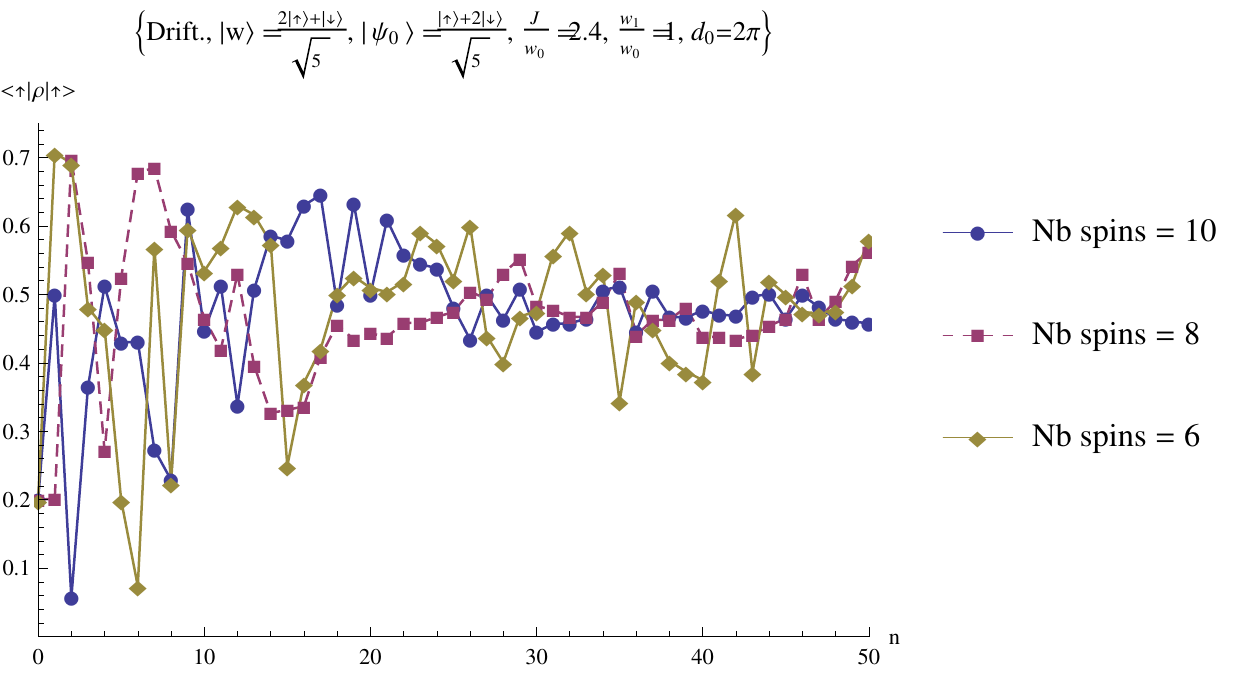}
\caption{\label{isingzplate} Evolutions of the coherence (up) and of the population (down) of the seventh spin of a ten spin chain coupled by a nearest-neighbor Ising-Z interaction. Each spin is submitted to a drift kick bath.}
\end{center}
\end{figure}

\begin{figure}
\begin{center}
\includegraphics[width=7.7cm]{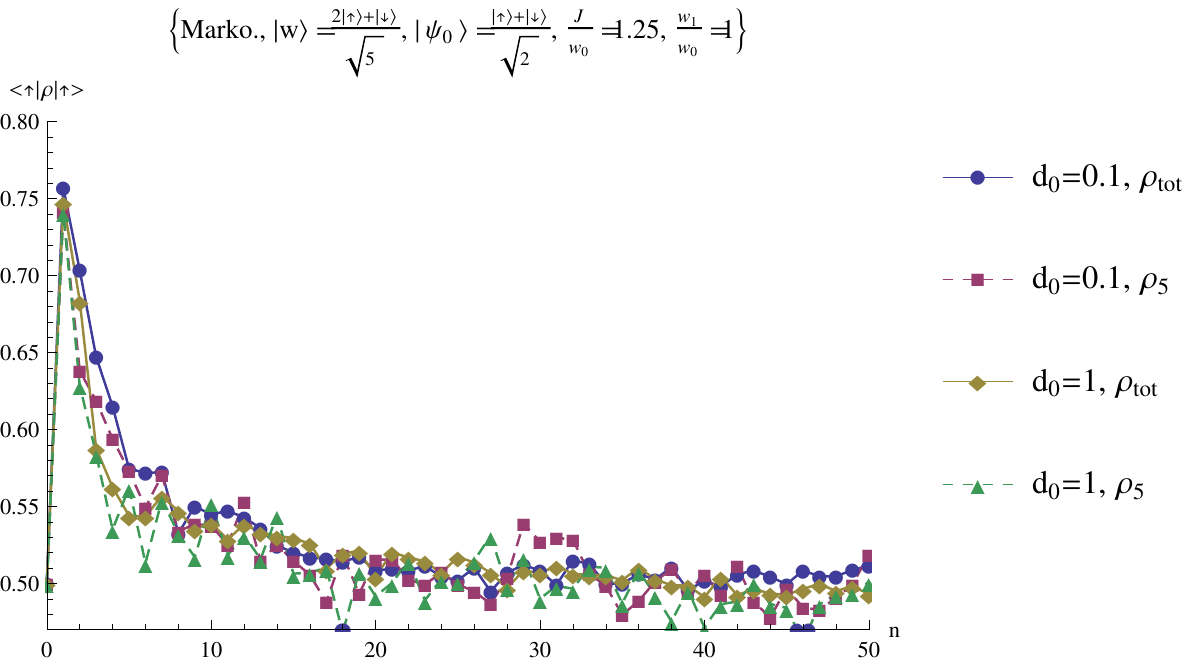}
\includegraphics[width=7.7cm]{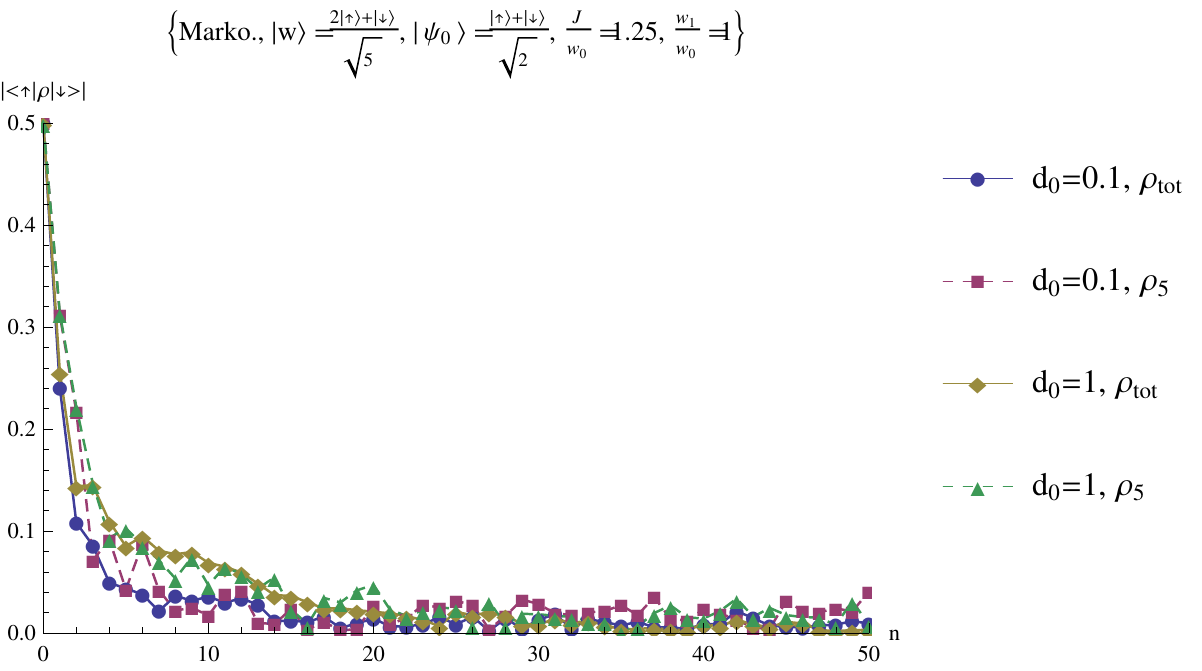}
\caption{\label{statisingxpopetcohetotavecd0} Evolutions of the population (up) and of the coherence (down) of a ten spin chain ($\rho_{tot}$) and of the fifth spin of the chain ($\rho_5$) with respect to $d_0$. Each spin is submitted to a Markovian kick bath and is coupled by the Ising-X interaction.}
\end{center}
\end{figure}

We have just seen the evolution of the disorder of a kicked spin chain. We are now interested by the effect of the disorder on the population and on the coherence of one spin of the chain and of an average spin. 

For the Heisenberg coupling, consider fig. \ref{popcoheheisenbergaveccouplage} and \ref{popcoheheisenbergsanscouplage}. They show a comparison between a coupled spin chain (fig. \ref{popcoheheisenbergaveccouplage}) and an ensemble of one thousand spins without coupling (fig. \ref{popcoheheisenbergsanscouplage}) with respect to an increase of the initial dispersion. We take a large number of spins in the ensemble because we know that from this number all spin ensembles have the same behaviors (see \cite{viennot2013}). The graphics show that, when there is a coupling, largest is the dispersion with a large coupling, more the population and the coherence of one spin and of an average spin of the chain go to the maximal lost of information : the microcanonic distribution. Each spin population follows the evolution of the population of an average spin of the chain. The observations are different for one thousand spins without coupling. The population of the ensemble follows a similar evolution than the one of the coupled chain but not one spin of the ensemble. The down graphic of fig. \ref{popcoheheisenbergsanscouplage} represents the coherence evolution of only one spin of the ensemble. It always oscillates and not follows the average evolution. So, the coupling induces that the population of one spin follows the behavior of the average spin of the chain. 

The analyses of the coherence also indicate another phenomenon. Largest is the initial dispersion, more the coherence falls to 0, what is not the case for the spin ensemble. The coherence goes to a value $\rho_{min} \geq 0$ (for more information, see \cite{viennot2013}). Fig. \ref{popcoheheisenbergavecetsanscouplage} shows the same thing than above, but with a comparison between the stationary and the microcanonical bath for a coupled chain, and a microcanonical bath for some ensembles characterized by various spin numbers. We see that, lower is the number of spins in the ensemble, more the population and the coherence oscillate. An ensemble of at least one hundred spins has nearly the same behavior than a chain of ten spins with a nearest-neighbor interaction. In addition the coherence of a chain submitted to a stationary kick bath has the same behavior than if the bath is microcanonical. The Heisenberg coupling allows to transmit the disorder to the spin chain but also to induce a larger disorder into the chain, what we have seen in the previous section.\\

Concerning the Ising-Z evolution, since there is a large entropy for $\frac{J}{w_0}$ not too small, the spin population quickly relaxes to the microcanonical distribution and the coherence rapidly falls to 0, for all the initial dispersions. This is as for the Heisenberg coupling when the dispersion is high. But, whereas for the Heisenberg coupling each spin follows the evolution of an average spin, this is not the case for the Ising-Z coupling (see fig. \ref{isingzvecteurpropre}).

A really important observation for this coupling is an initial small "plateau", as we can see on fig. \ref{isingzplate}. This one is visible for the individual spin coherence but not for the average coherence because of the added oscillations of each spin. The coherence plateau does not depend on the dynamics, on the initial dispersion and apparently on the spin number. It does not correspond to a maximal coherence and its value is about 0.2-0.3. The Ising-Z plateau more looks like to a low decreasing of the decoherence than a real plateau. But during this plateau, there is an evolution of the population of each spin. The spin population can oscillate before the fall of the coherence to $0$. The little plateau is due to the spin interaction which tends to keep all spins in the same $z$ projection. One spin try to remain coherent with their neighbors.

Finally, for the Ising-X model, as the Heisenberg coupling, each spin follows the average evolution (see fig. \ref{statisingxpopetcohetotavecd0}). If $d_0$ is large or really small, there is a fall of the coherence to 0 and a relaxation of the population to the microcanonical distribution (fig. \ref{statisingxpopetcohetotavecd0}). This completes the previous observations. No matter the distribution of the bath and the dispersion, there is always a large disorder into the spin chain, due to the interaction. The disorder seems to be maximal.\\

For all the spin chain couplings, if the interaction decreases, there is a mix between the evolution without interaction (in a first part) and one due to the coupling (the second part). The first part of the evolution will only depend on the initial dispersion $d_0$ and the kick bath chosen (see \cite{viennot2013}).

\subsection{Analysis of the behavior with the Husimi distribution}
The aim of this part is to understand the results obtained in the two previous part. For this, it is interesting to analyze the Husimi distribution and so to use a classical analysis. The Husimi distribution is defined by 
\begin{equation}
\label{husimi}
H(\theta,\varphi) = |<\theta, \varphi|\psi>|^2
\end{equation}
with $|\theta, \varphi> = \cos(\frac{\theta}{2}) |\uparrow> + e^{\i \varphi} \sin(\frac{\theta}{2}) |\downarrow >$. The Husimi distribution measures the quasiprobability distribution of a quantum state onto the classical phase space (here, the sphere of the classical spin direction). In other words, the Husimi distribution gives the most probable classical state of a spin. The sphere of the phase space will be represented by an azimuthal projection map (north pole at the center and south pole as being the limit circle). The entanglement processes is also shown by the Husimi distribution which becomes uniform for a maximal entanglement state.

\subsubsection{Heisenberg coupling}

\begin{figure}
\begin{center}
\includegraphics[width=7.7cm]{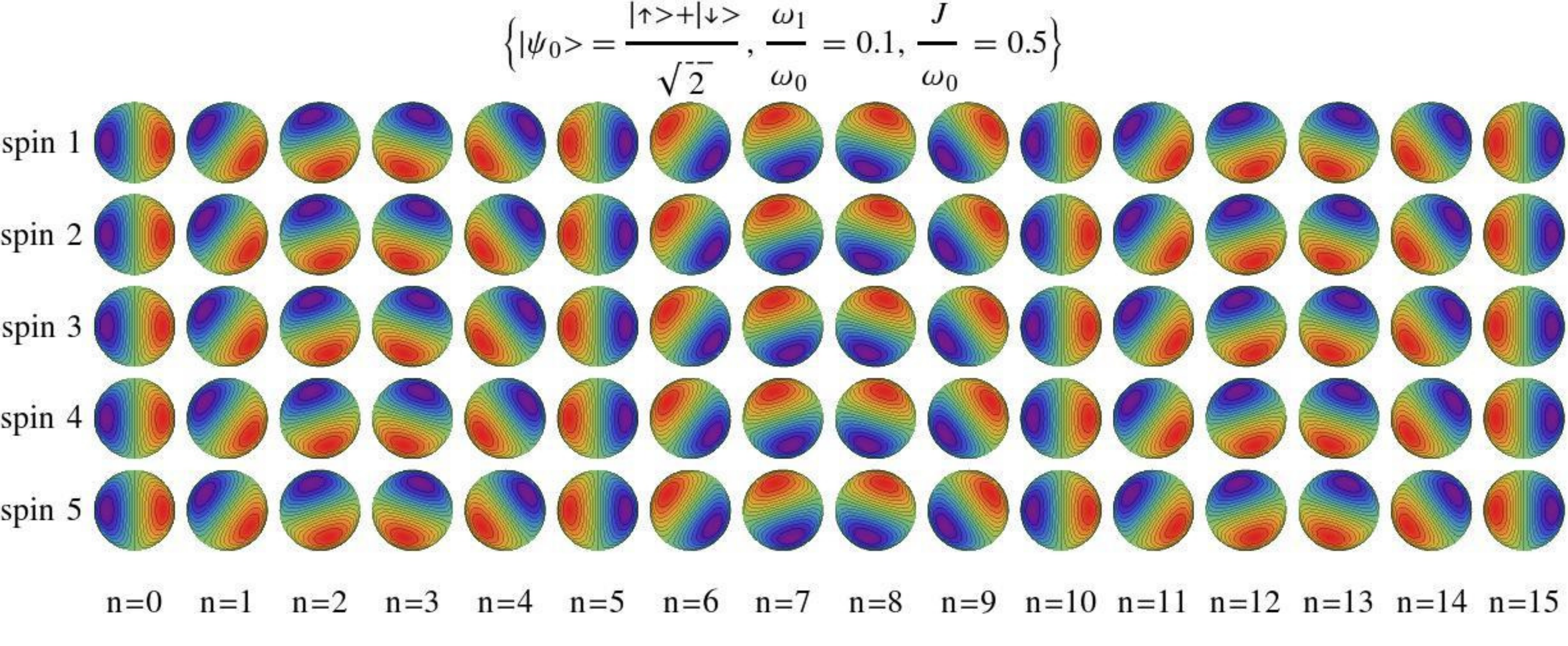}
\includegraphics[width=7.7cm]{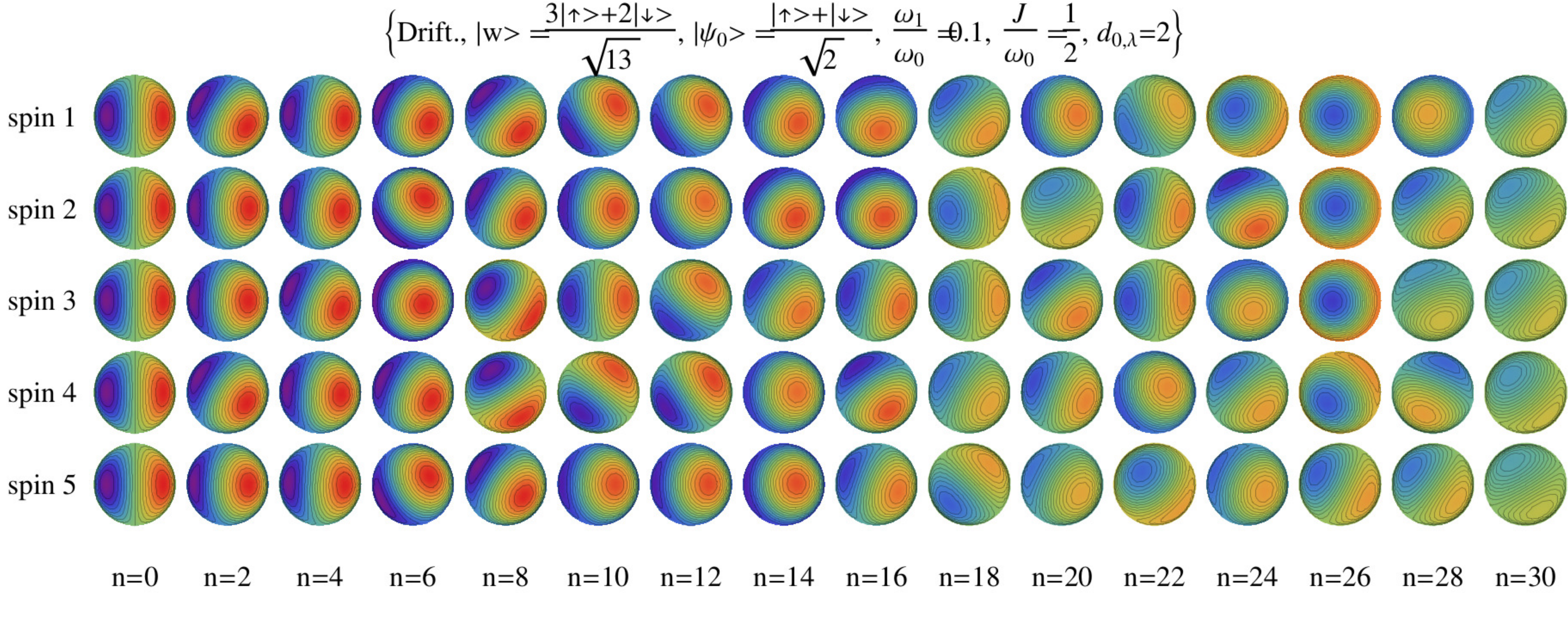}
\caption{\label{precession} Evolution of the Husimi distribution of the five spins of a chain coupled by a nearest-neighbor Heisenberg interaction without kick (up) and submitted to a drift kick bath (down). The highest probability is represented in red and the smallest one is represented in blue.  The entanglement process is also shown by the Husimi distribution. In this case, the spheres go to the green color.}
\end{center}
\end{figure}

\begin{figure}
\begin{center}
\includegraphics[width=7.7cm]{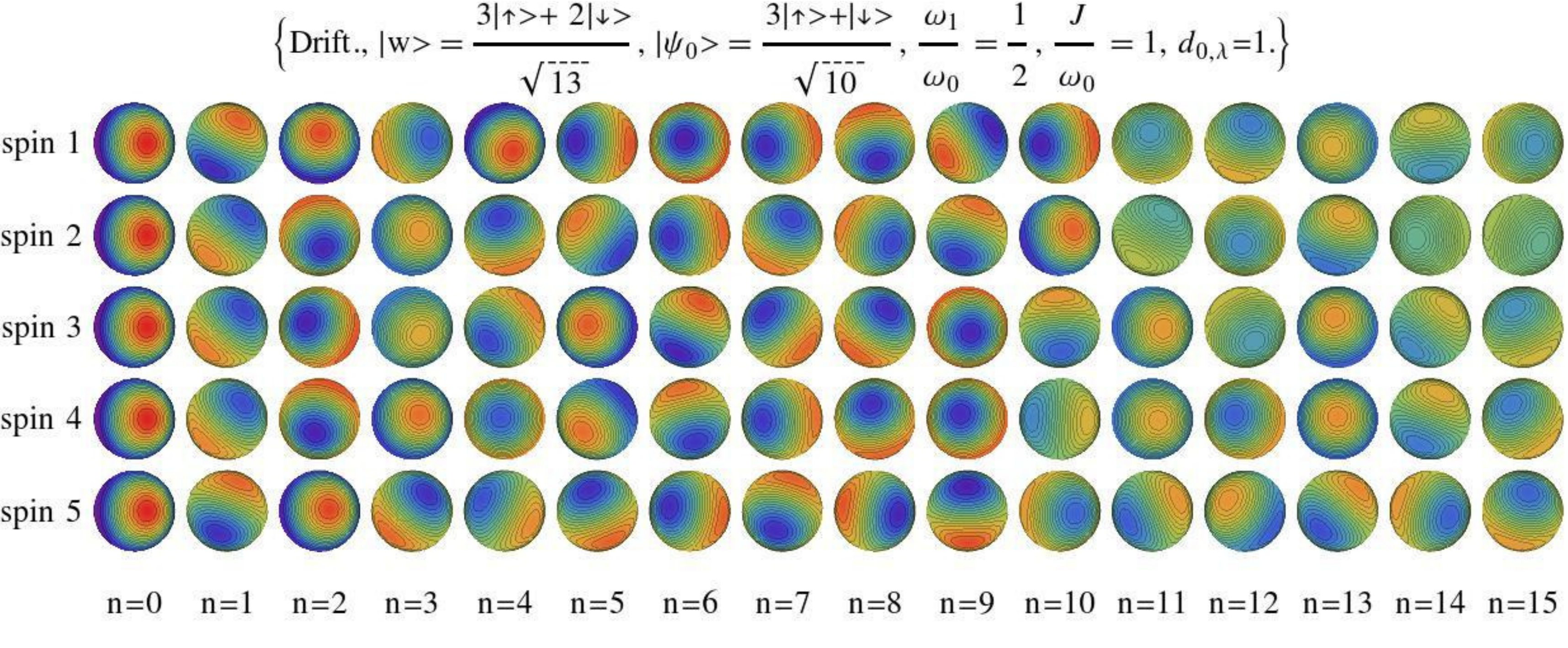}
\caption{\label{husimiisingZ} Evolution of the Husimi distribution of each spin of a chain coupled by a nearest-neighbor Ising-Z interaction and submitted to a drift kick bath. The highest probability is represented in red and the smallest one is represented in blue.  The entanglement process is also shown by the Husimi distribution. In this case, the spheres go to the green color.}
\end{center}
\end{figure} 

We begin with a spin chain coupled by a Heisenberg interaction. Second graphic on fig. \ref{precession} shows a general behavior of the chain for a medium initial dispersion. At the beginning, all spins are in the same direction for each kick. From the kick number three, the spins begin to be orientated in various directions. The disorder appears into the spin chain. Larger is the kick number, more the colors mitigate. The spins begin to be entanglement with their neighbors. No particular phenomenon is observed for the edge spins. 

There is also another effect that we cannot see in the population or in the coherence graphics. The magnetic Zeeman field $\vec{B}$ induces a spin precession around the $z$ axis which can be viewed by using the Husimi distribution (see first graphic on fig. \ref{precession}).

So, the Heisenberg interaction tends to align the spins in the same spatial direction. The coupling is isotropic. If all spins are in the same initial state, for a small initial dispersion of the control parameter ($d_0 \approx 0$), they are kicked in the same direction with the same strength and the same delay. The coupling does not modify the spin orientation, for each kick they are in the same direction. The entropy and the entanglement remains really small (fig. \ref{entropyheisenbergd0}). But, for a large initial dispersion, each spin is kicked with different strengths and delays. So the evolution and the direction of one spin after a kick, is different from the other spins. But, the interaction still tends to align all spins in the same direction. In order to obtain this result, each spin of the chain gets entangled with their neighbors (which is seen on fig \ref{entropyheisenbergd0}). These two effects are in opposition and the spins disturb their neighbors during their oscillations. They lose their initial movements and go to the maximum lost of information : the microcanonical distribution. 

\subsubsection{Ising-Z coupling}
Fig. \ref{husimiisingZ} shows the spin chain evolution. The little plateau seen in sec. 3.2 can be viewed here from the second kick to the nineth one. The use of the Ising-Z interaction for the spin chain generally induces a fall of the coherence after the little plateau to 0 and a relaxation of the population to $0.5$ (if $|w \rangle \neq \frac{1}{\sqrt{2}} (|\uparrow \rangle + |\downarrow \rangle)$, $|\uparrow \rangle$ or $|\downarrow \rangle$, see after). To understand this phenomenon, it is necessary to take into account the precession, the Ising-Z coupling and the kicks. The no-kicked spins remain at their initial positions with a precession movement around the $z$-axis (like with the Heisenberg coupling on the first graph of fig. \ref{precession}). They do not get entangled together except if they are initially on various states. More the states between two spins are different, more they get entangled.

The spins are submitted to a Zeeman field and their own fields. They are both in the $z$ direction and induce only a precession around the $z$-axis. But if the spins are in different initial states, they do not have the same precession. Because of the coupling, one spin gets entangled with their nearest-neighbors. This entanglement tries to induce a same precession time for the considered spin and its neighbors. More the speed precession between two spins is different, more their entanglement is large.

Now, if the chain of spins is kicked, two phenomena appear. Each kick modifies the spin orientations. So each spin has a speed precession modified by their neighbors. This requires an entanglement and so a decoherence process. The second one concerns the edge spins. When the kicks modify the spin orientations, the spins could begin their rotation toward both opposite directions (for example, if the spin direction is $x$ and the kick direction is $z$, they could begin their rotation either toward $y$ or $-y$). Since the edge spins are less influenced than the others (they have only one neighbor), they can not have a precession in the same direction than the others. The spins which are next to them, are attracted in two different directions. So they go in a different direction than the edge spins and the others spins next to them. After, they transmit this disorder to the other spins. Finally, all spins are in a different orientation and the entanglement is large. 

\subsubsection{Ising-X coupling}
\begin{figure}
\begin{center}
\includegraphics[width=7.7cm]{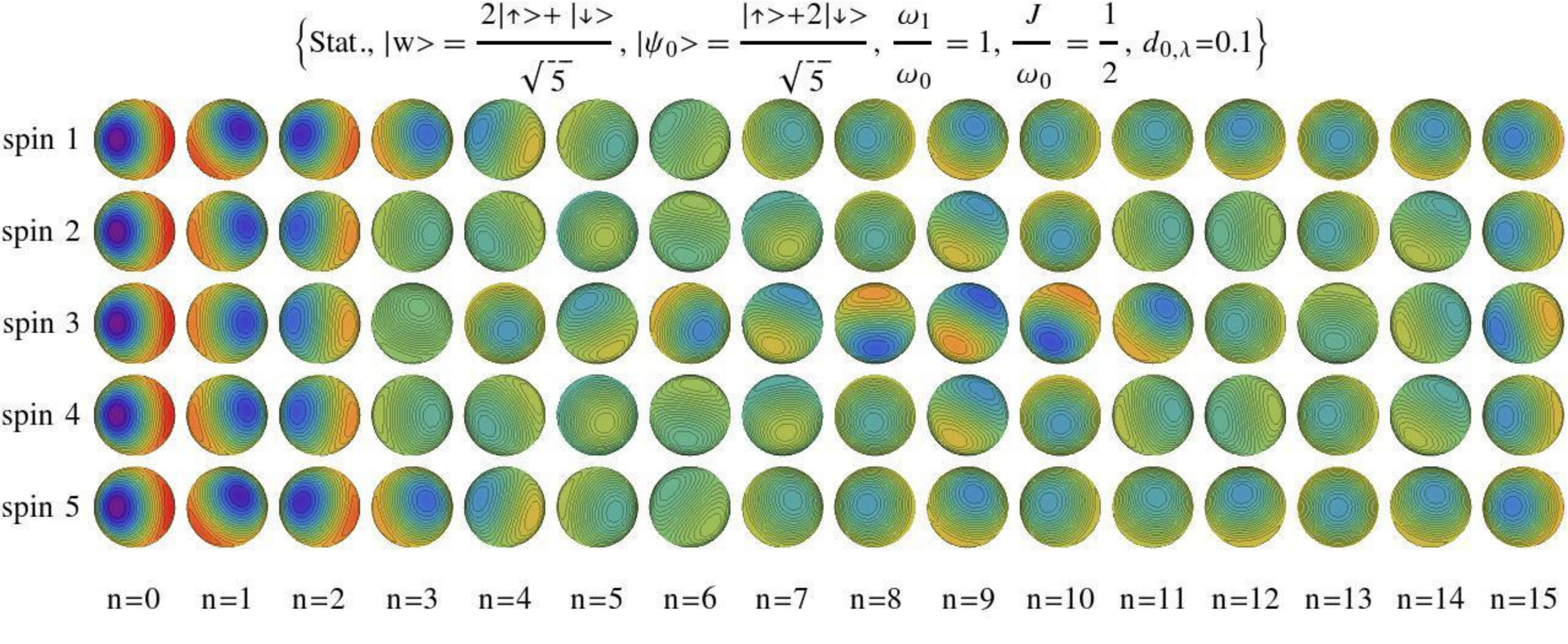}
\caption{\label{husimiisingxcoheedgespin} Evolution of the Husimi distribution of each spin of a five spin chain coupled by an Ising-X interaction. Each spin is submitted to a drift kick bath with a small initial dispersion of the kick strengths and the kick delays. The highest probability is represented in red and the smallest one in blue. The entanglement process is also shown by the Husimi distribution. In this case, the spheres go to the green color.}
\end{center}
\end{figure}

The Ising-X coupling rapidly induces a microcanonical behavior into the chain with a lot of disorder. This can be better seen on fig. \ref{husimiisingxcoheedgespin}.

\begin{figure}
\begin{center}
\includegraphics[width=9.cm]{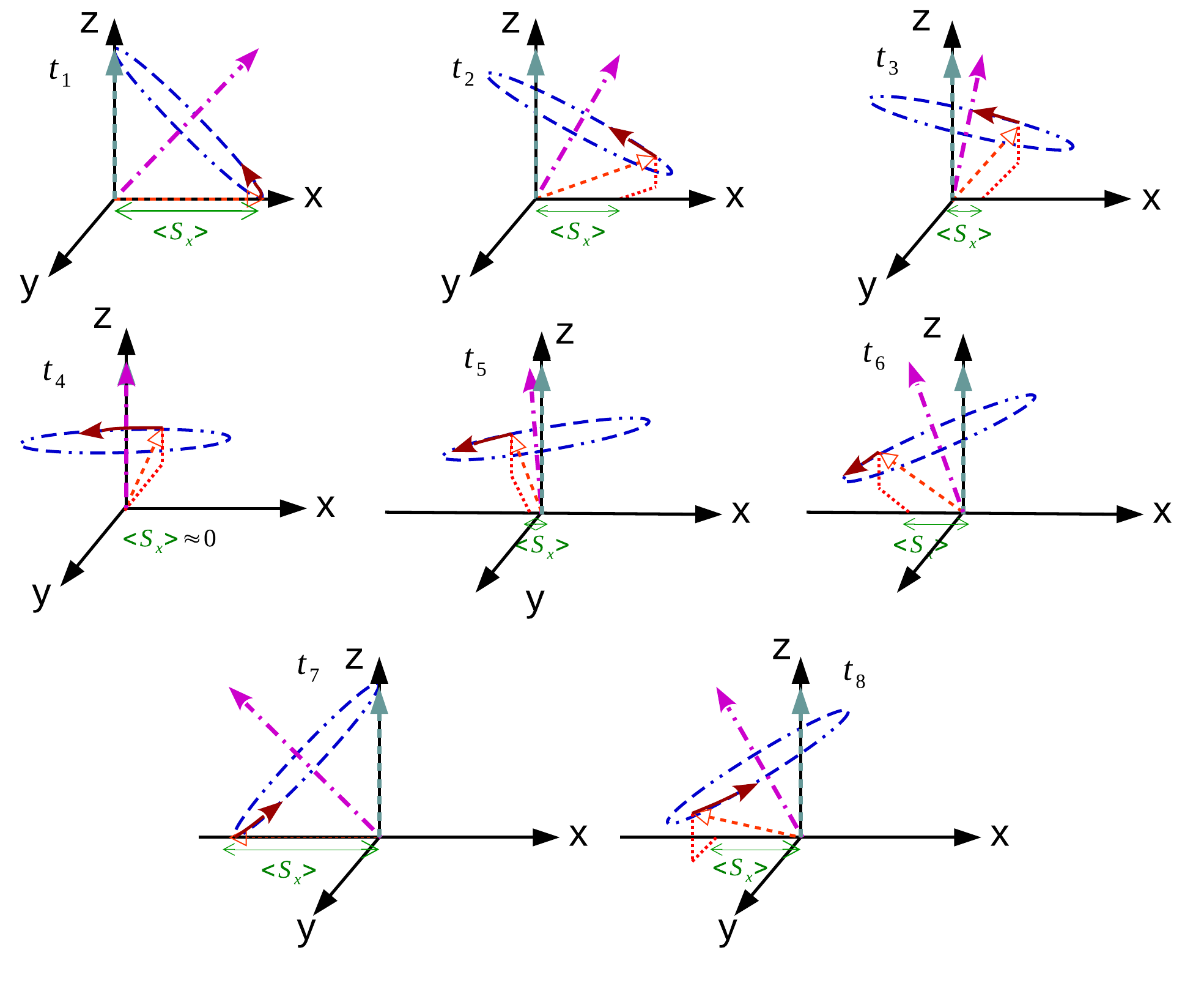}
\caption{\label{isingxevolutionschematic} Schematic representation of the evolution of one spin of the chain (red long-dashed arrow) induced by the total field (magenta dash-dotted arrow). The total field is composed by the Zeeman magnetic field (cyan arrow) and by the field in the $x$ direction (green) induced by the neighbor spins. This last field is characterized by the projection on the $x$ direction, of the average of the spin operator $S$ of its neighbors (called $\langle S_x \rangle$). The blue dash-dotted circle is the representation of the precession with respect to the spin position. The considered spin follows the precession circle in the direction indicated by the red full arrow. We suppose that all the spins are initially in the $x$ direction and that the Zeeman magnetic field has the same intensity than the local field.}
\end{center}
\end{figure}

\begin{figure}
\begin{center}
\includegraphics[width=7.7cm]{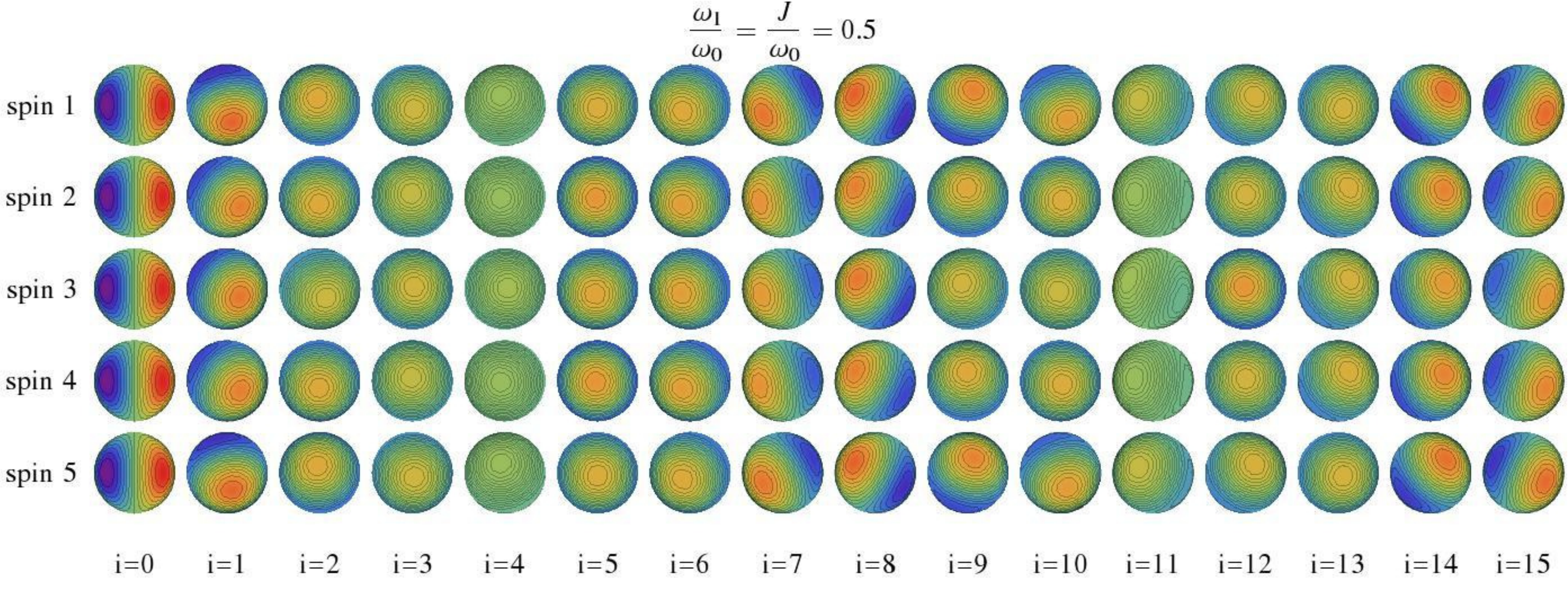}
\caption{\label{isingxmovement} Evolutions of the Husimi distribution of the spin chain submitted to the Ising-X interaction without kick. Each spin is in the initial state $\psi_0= \frac{1}{\sqrt{2}}(|\uparrow \rangle + |\downarrow \rangle)$. The highest probability is represented in red and the smallest one in blue.  The entanglement process is also shown by the Husimi distribution. In this case, the spheres go to the green color.}
\end{center}
\end{figure}

In order to understand more the Ising-X coupling  model, it is necessary to see the spin evolution without kick. We consider a coupled spin chain without kick. In this model a spin feels two magnetic fields. The first one is the Zeeman magnetic field which is in the $z$ direction and allows an energy level splitting. The second one is given by the term $-J S_{x_1}.\langle S_{x_2} \rangle$ and could be interpreted as a field in the $x$ direction induced by the neighbor spins. These two components generate a total field which is the precession direction of the spin. But this total field is constantly modified. To understand this, consider the following example and fig. \ref{isingxevolutionschematic}. At time $t_1 =0$, the spins are in the $x$ direction. So the total field is at $\gamma = \frac{\pi}{4}$ ($\gamma$ is the angle between the $x$ direction and the total field in the $(x,z)$ plane. For the example, we have chosen the Zeeman magnetic field with the same intensity that the local field). The spins begin their precession around this new axis. The radius of the precession circle is the distance between the spin position and the total field direction. Since the spins move, the total field of each spin also moves (induced by the neighbor spins, but here, since all spins are in the same initial state, the total field is the same for each spin and then their displacements are the same) and goes toward $z$ with a decrease of $\langle S_x \rangle$ (see $t_2$ and $t_3$ of fig. \ref{isingxevolutionschematic}). When the total field direction is along the $z$ axis, spins turn around this axis and $\langle S_x \rangle$ becomes zero ($t_4$ of fig. \ref{isingxevolutionschematic}). The spins follow their precession around $z$ and quickly, $\langle S_x \rangle$ becomes different from zero ($t_5$ of fig. \ref{isingxevolutionschematic}). The total field is different from the $z$ axis and constantly goes to the $x$ axis in the direction $-x$ ($t_6$ of fig. \ref{isingxevolutionschematic}). When they arrive to the $x$ axis, the spins go to the opposite direction (after an half precession around the $x$ axis) toward the $z$ axis ($t_7$ of fig. \ref{isingxevolutionschematic}), and so on. When one spin modifies the precession direction of its neighbors, this induces an entanglement between them. This can be seen fig. \ref{isingxmovement}. In this graphics we see the Husimi distribution of a chain of five spins. We clearly see that they begin their movement from the $x$ direction to the $-x$ one. During this movement the fading of the red and blue colors shows a spin entanglement. 

When the spins of the chain are submitted to kicks, their directions are also modified. The spin direction modifications are induced by the spin precession but also by the kicks. Thus, spins always become entangled.

For an initial dispersion of the control parameters, the kicks move each spin in a different direction from its neighbors. So a spin turns around a total field given by the Zeeman magnetic field and two others components which are induced by the neighbors of the considered spin. The total field of one spin is different from the one of another spin. So this produces a large entanglement between the spins which is conserved.\\

In brief, the Heisenberg coupling is an isotropic coupling. It induces a same orientation for two coupled spin. If the strengths and the delays are the same for all spins, they are in the same direction for each kick. The coupling has no effect. However, if there is an initial dispersion of the kicks, all spins are differently kicked. The disorder into the classical bath is transmitted to the chain. In addition, the coupling induces an entanglement between the spins which transmits the disorder from one spin to another one. Thus the disorder is transmitted along the spin chain. The coupling also allows that one spin has the same behaviors than the average. In spite of the disorder induced by the entanglement, the coupling conserves the spin ensemble evolution

The Ising-Z and Ising-X coupling induces a lot of disorder and entanglement. The interaction parameter is a source of the entanglement and this one and the initial dispersion of the kick bath, of the disorder. Thus these parameters allow a large transmission of the disorder from the bath to the spin chain and into the spin chain. For Ising-Z coupling, a little "plateau" allows the conservation of the coherence during a small number of kicks with some large oscillations. During this kick number, the population of each spin can oscillate.\\

We have just seen that all spin interactions and an initial dispersion of the kicks induce a lot of disorder into the spin chain with a large decoherence and a population relaxation. Is that possible to avoid the disorder appearance into the spin chain?

\section{Some conservation of the order into the spin chain}
Now that we have seen the general behavior of a chain, we are interested by the possibility of keeping some order into the chain in spite of the disorder of the bath.

\subsection{Kicks in an eigenvector direction}
\begin{figure}
\begin{center}
\includegraphics[width=7.7cm]{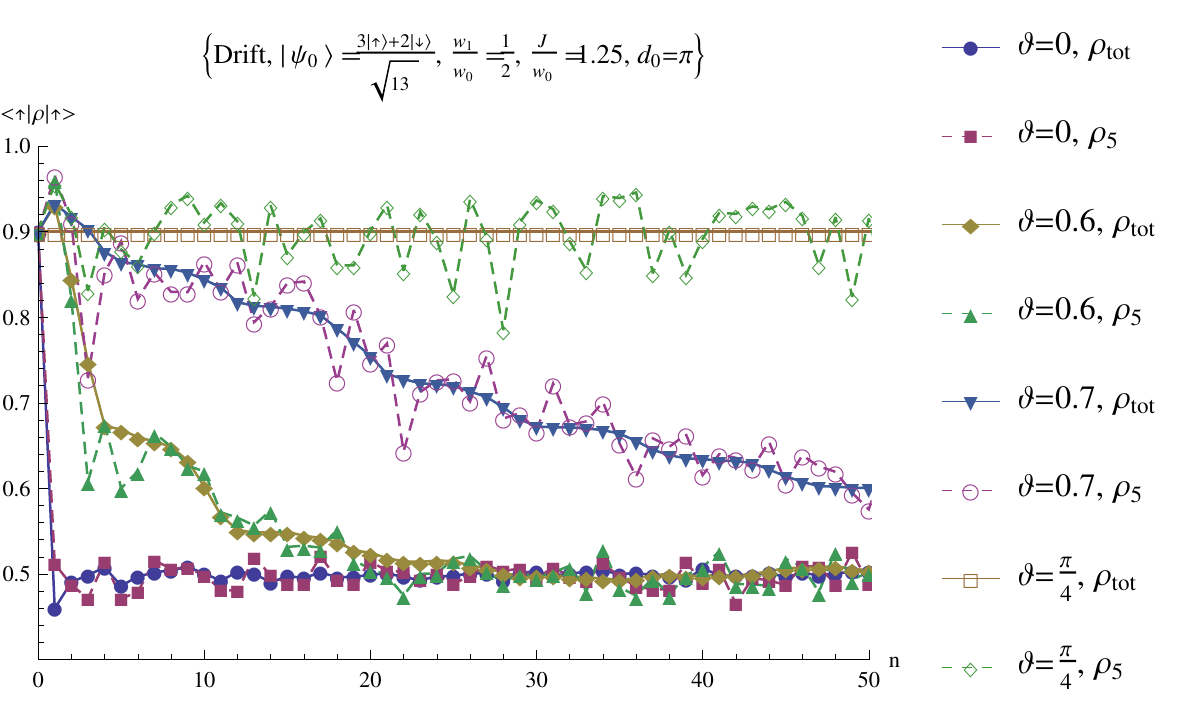}
\includegraphics[width=7.7cm]{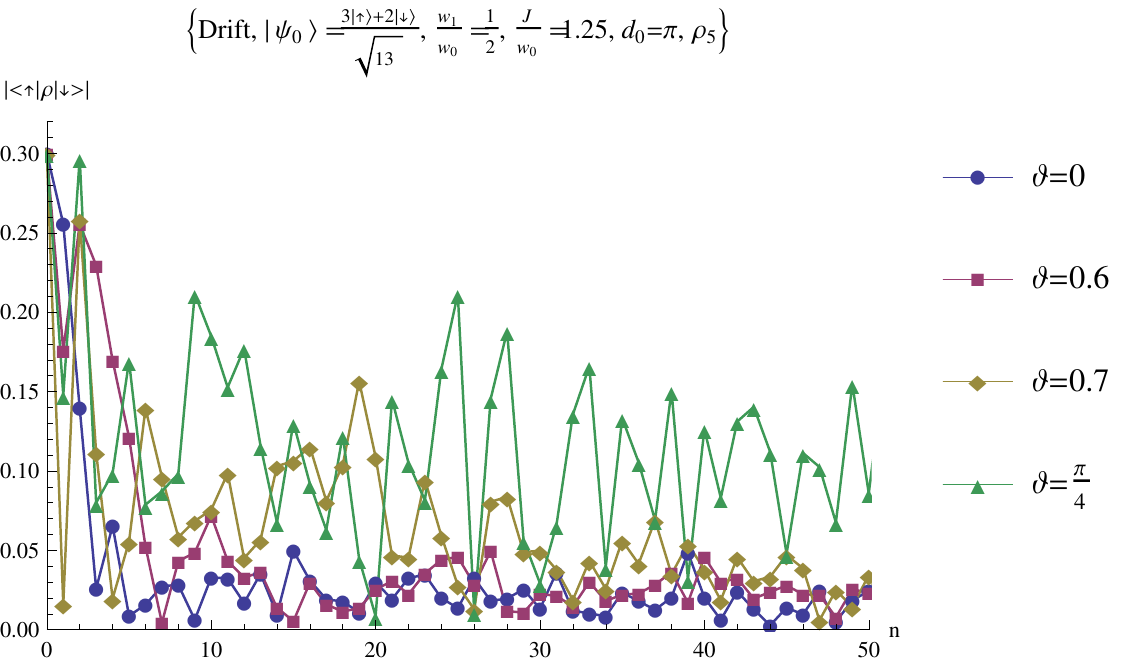}
\caption{\label{heisenbergdriftwiththeta} Evolution of the population (up) and of the coherence (down) of a ten spin chain ($\rho_{tot}$) and of the fifth spin of the chain ($\rho_5$) with an increase of $\vartheta$ ($|w> = \cos(\frac{\pi}{4} - \vartheta) |\uparrow > + \sin(\frac{\pi}{4} - \vartheta) |\downarrow >$). Each spin is submitted to a drift classical kick bath and coupled by the Heisenberg interaction .}
\end{center}
\end{figure}

\begin{figure}
\begin{center}
\includegraphics[width=7.7cm]{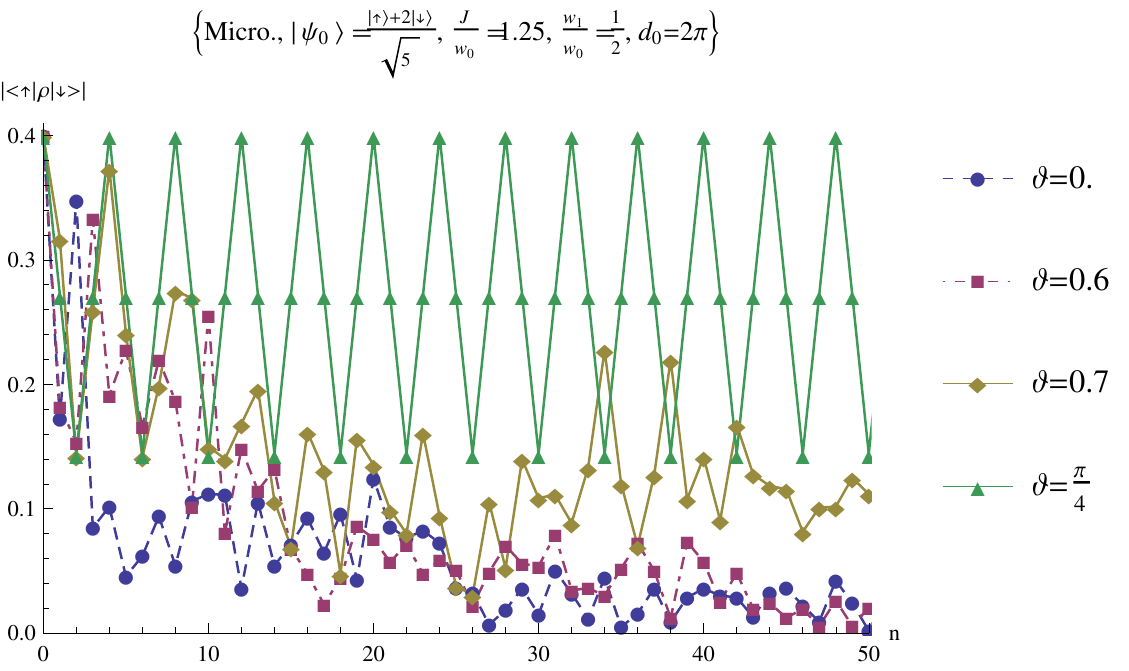}
\caption{\label{isingzvariationw} Evolution of the fifth spin of a ten spin chain coupled by the Ising-Z interaction. The kick direction is characterized by $|w\rangle = \cos(\frac{\pi}{4}-\vartheta) |\uparrow \rangle + \sin(\frac{\pi}{4} - \vartheta) |\downarrow \rangle$). Each spin of the chain is submitted to a microcanonical kick bath.}
\end{center}
\end{figure}

\begin{figure}
\begin{center}
\includegraphics[width=7.7cm]{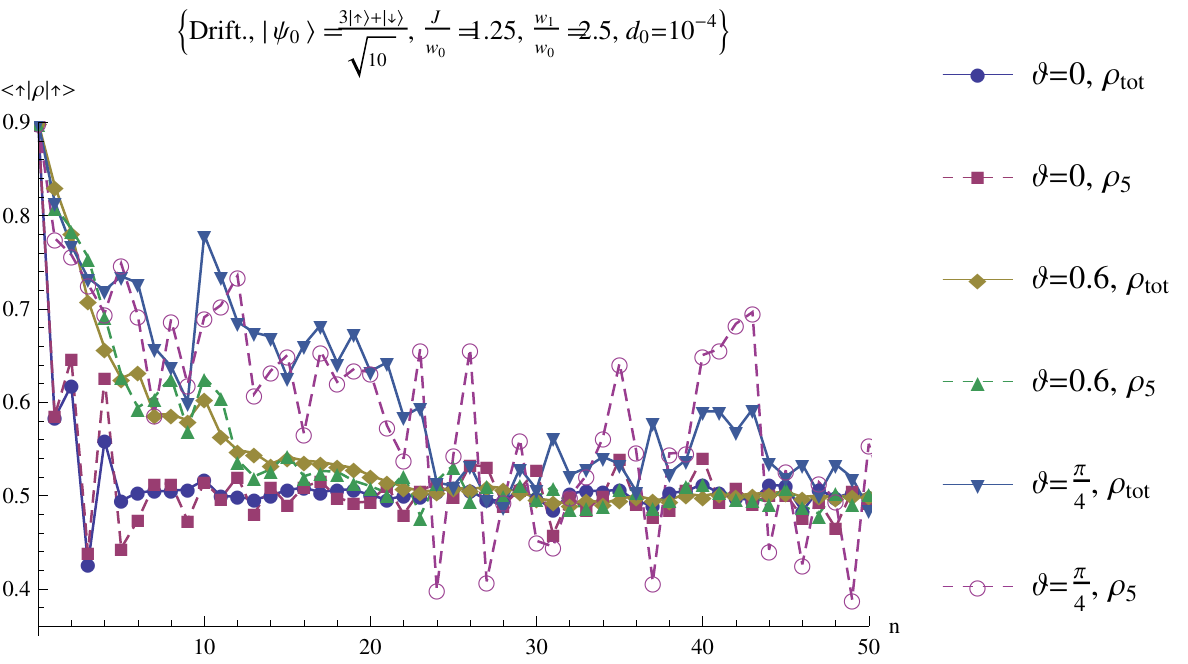}
\includegraphics[width=7.7cm]{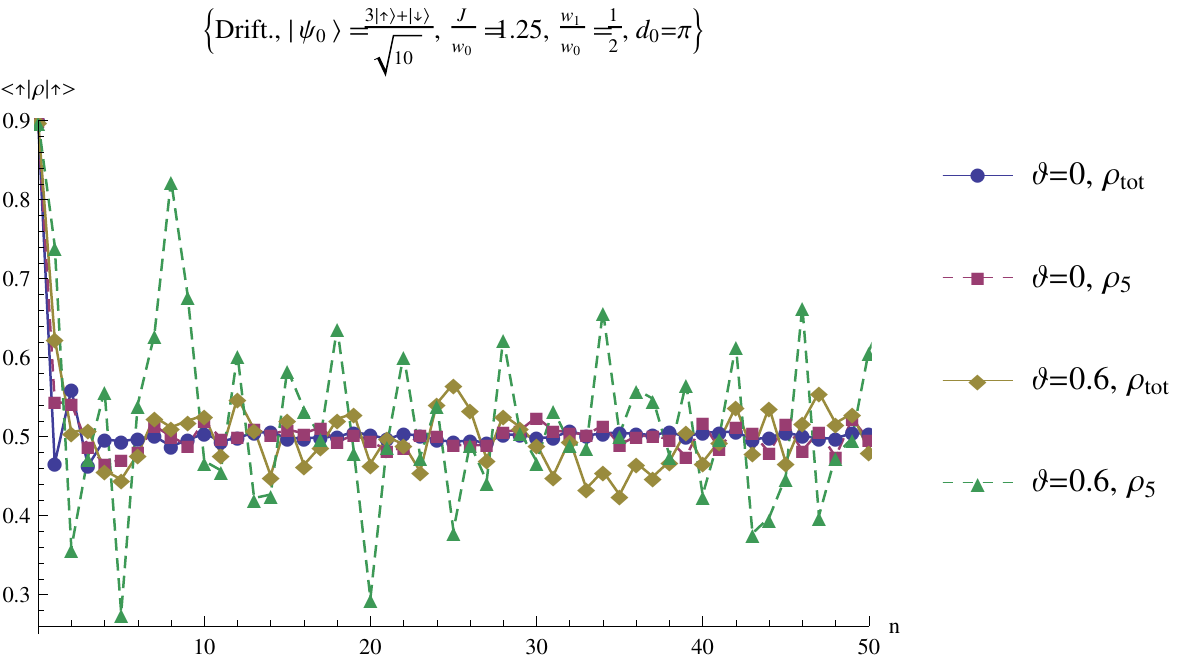}
\caption{\label{isingxvitesserelaxation} Evolution of the population of a ten spin chain ($\rho_{tot}$) and of the fifth spin of a chain ($\rho_5$). The kick direction is characterized by $|w \rangle = \cos(\frac{\pi}{4} - \vartheta) |\uparrow \rangle + \sin(\frac{\pi}{4} - \vartheta) |\downarrow \rangle $. This chain is submitted to a drift kick bath and is coupled by the Ising-X interaction. The up graphic is for a small initial dispersion of the kick strengths and of the kick delays, and the down one is for a large initial dispersion.}
\end{center}
\end{figure}

In the Heisenberg or in the Ising-Z coupling we see on fig. \ref{heisenbergdriftwiththeta} and \ref{isingzvecteurpropre} that more the kick direction is close to an eigenvector ($|\uparrow >$ or $| \downarrow >$) less the population rapidly decreases to a microcanonical evolution. For the Heisenberg coupling, the population of each spin goes to the average spin population whereas for the Ising-Z coupling, the population of each spin goes to their own initial population. The fact that, if $|w \rangle \to |\uparrow \rangle$ the spin populations go to there own initial state or to the spin average state and not relax to $0.5$, is because the eigenvectors are the same than the Heisenberg one and than without coupling. The explanation is the following. If $|w \rangle = |\uparrow \rangle$ or $|\downarrow>$, $(|\uparrow \rangle ,|\downarrow>)$ are also eigenvectors of the kick operator $W$. So they are eigenvectors of the monodromy operator eq. \ref{monodromy}. The dynamics only induces a phase between the both components of the spin wave function. The phase does not act on the population but only on the coherence. 

The population evolution in the case of the Heisenberg coupling is similar to the one observed for a spin ensemble \cite{viennot2013}. In spite of the disorder induced by the entanglement, the coupling keeps the spin ensemble evolution.

Concerning the decoherence process, the evolution is different if we consider an average spin of the chain or only one spin of the chain. Fig. \ref{heisenbergdriftwiththeta} and \ref{isingzvariationw} show that the fall of the coherence is more efficient for only one spin of the chain if $|w \rangle \to \frac{1}{\sqrt{2}}(|\uparrow \rangle + |\downarrow \rangle)$. In addition, this figure shows that if $|w \rangle = |\uparrow\rangle$ or $|\downarrow \rangle$, the coherence of each spin strongly oscillates. For a Heisenberg and an Ising-Z coupling, the average spin has a coherence which falls to 0 (with little oscillations) because of the added oscillations of each spin. For the Ising-Z coupling, the oscillation amplitudes decrease when $|\psi\rangle \to |\uparrow\rangle$ or $| \downarrow \rangle$

For the Ising-X coupling, this phenomenon is not present as we can see fig. \ref{isingxvitesserelaxation}. This can be understood by the fact that this coupling induces other eigenvectors of $H$. Either there is a fall of the coherence to 0 and a fast population relaxation, or if the dispersion parameter is really small, we can see a small decrease of the relaxation when the kick is similar to an eigenvector direction.

\subsection{Small spin Rabi frequency}
\begin{figure}
\begin{center}
\includegraphics[width=7.7cm]{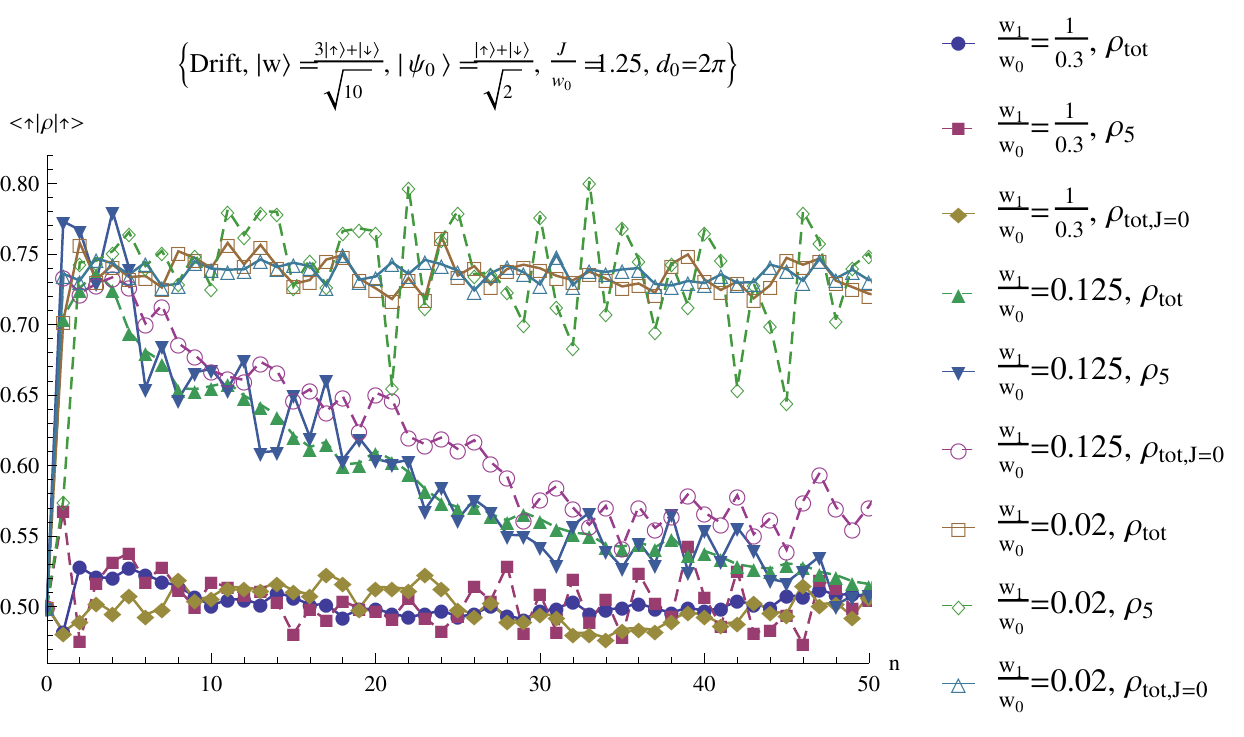}\\
\includegraphics[width=7.7cm]{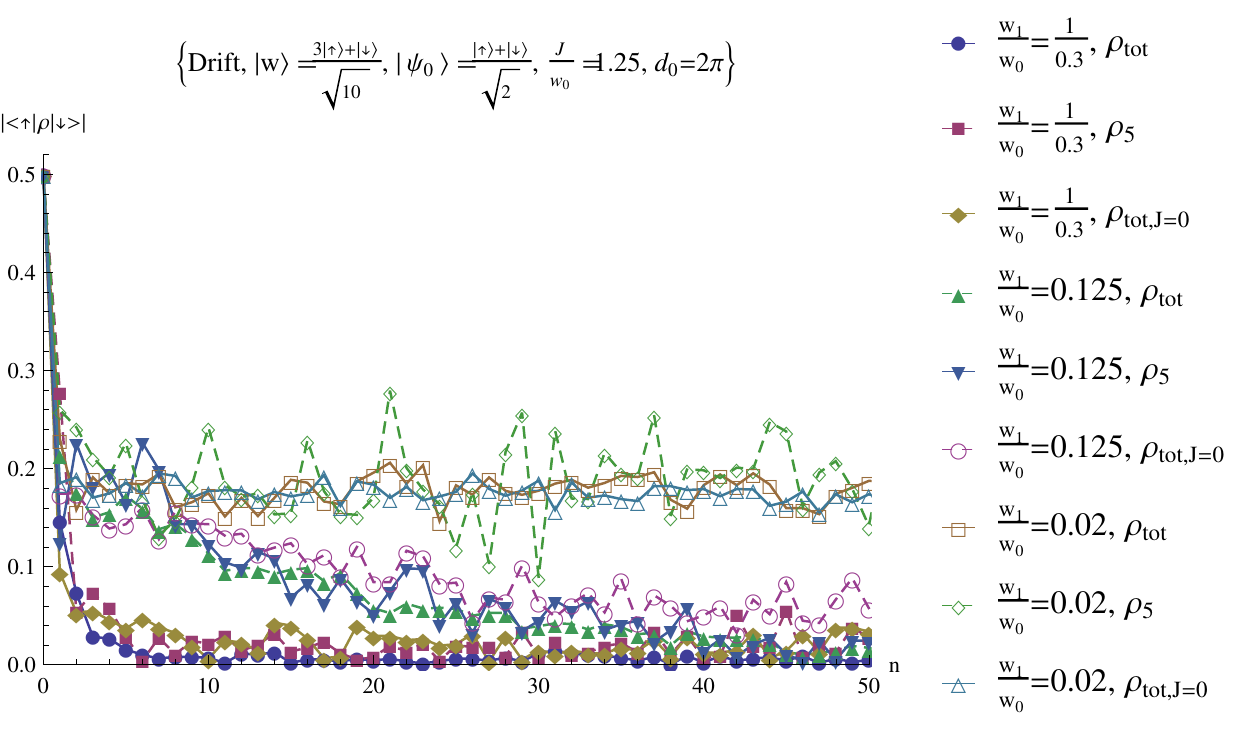}
\caption{\label{heisenbergdriftwithJ} Evolution of the population (up) and of the coherence (down) of a ten spin chain ($\rho_{tot}$), of one spin of the chain ($\rho_5$) and of an ensemble of one hundred spins ($\rho_{tot, J=0}$) with respect to an increase of $\frac{w_1}{w_0}$. Each spin is submitted to a drift classical kick bath. The spins of the chain are coupled by the Heisenberg interaction }
\end{center}
\end{figure}

\begin{figure}
\begin{center}
\includegraphics[width=7.7cm]{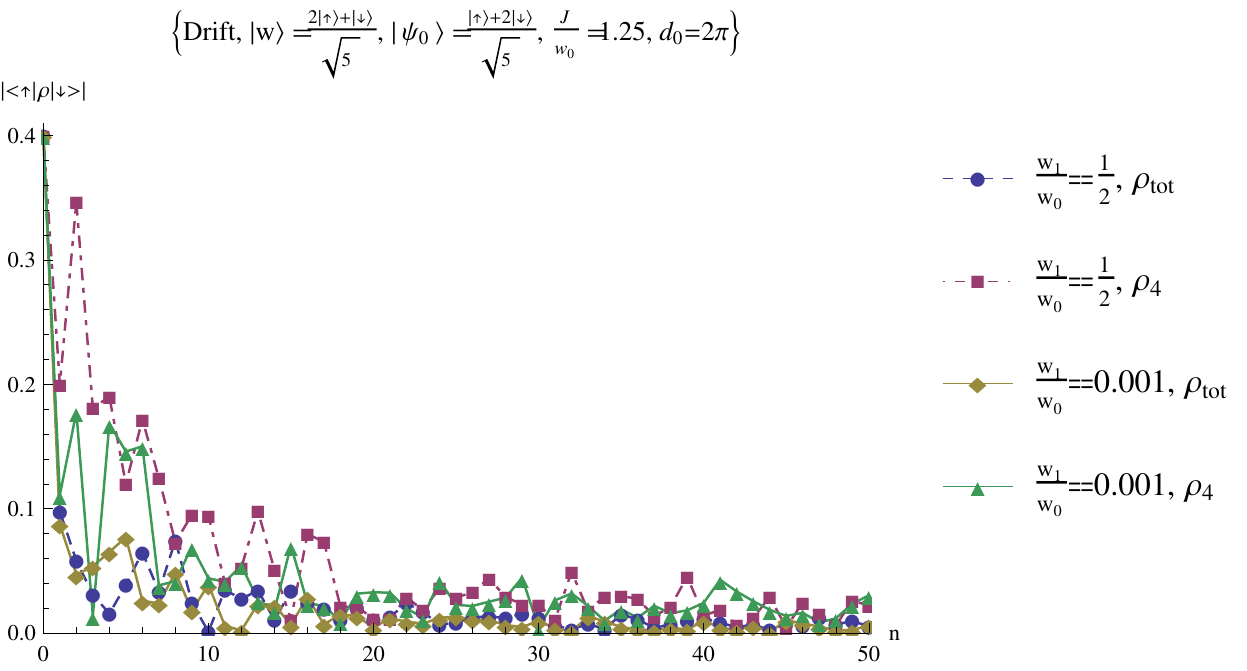}
\caption{\label{isingzevolutionwithd1} Evolution of the population (up) and of the coherence (down) of a ten spin chain ($\rho_{tot}$) and of the fourth spin of the chain ($\rho_4$). Each spin of the chain is submitted to a classical drift kick bath and is coupled by the Ising-Z interaction.}
\end{center}
\end{figure}

\begin{figure}
\begin{center}
\includegraphics[width=7.7cm]{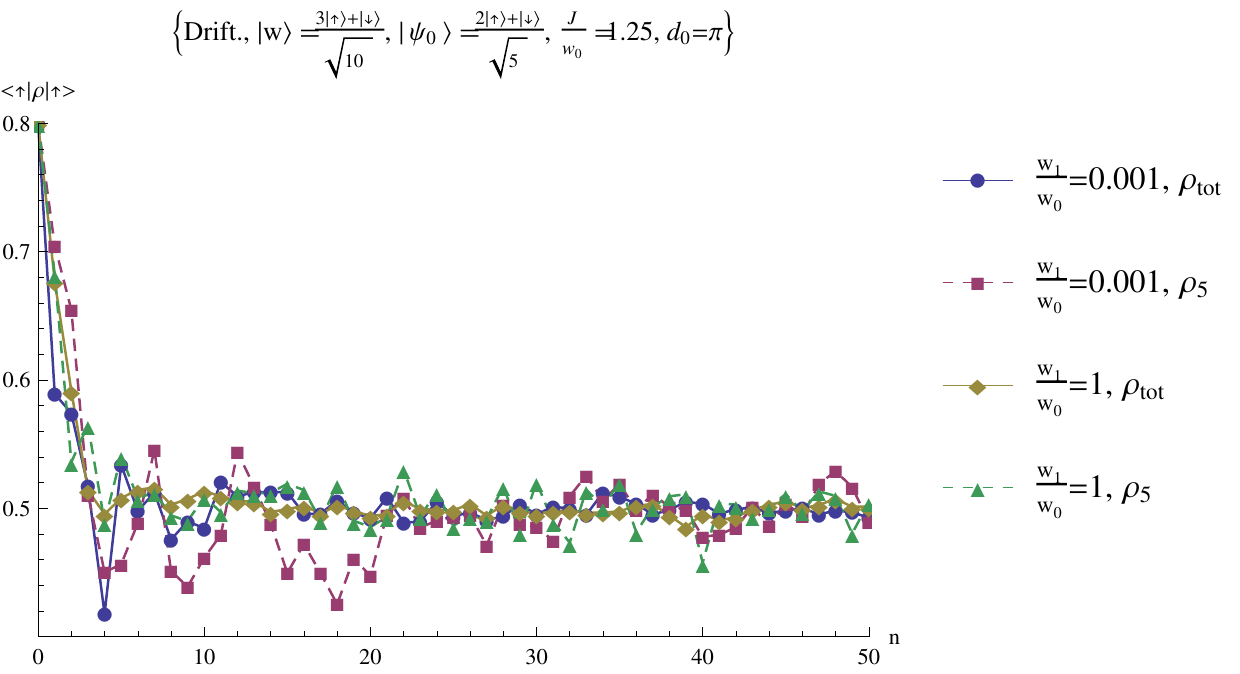}
\caption{\label{driftisingxpopavecd1} Evolution of the population of a ten spin chain ($\rho_{tot}$) and of the fifth spin of the chain ($\rho_5$) submitted to a drift kick bath with respect to $\frac{w_1}{w_0}$. Each spin is coupled to its neighbors by an Ising X interaction}
\end{center}
\end{figure}

Another important behavior is shown for a Heisenberg coupling on fig. \ref{heisenbergdriftwithJ}. This figure corresponds to a comparison between the population and the coherence evolution of a coupled spin chain, of one spin of the chain and of a spin ensemble with respect to $\frac{w_1}{w_0}$. For these graphics, we still see that one spin of the chain follows the average chain evolution. But also that if $\frac{w_1}{w_0}$ becomes small, the population less rapidly decreases to the microcanonical evolution and the decoherence decreases less faster (the first decrease is due to the initial dispersion). This can be understood by the fact that if $\frac{w_1}{w_0}<<1 $, $w_0 >>w_1$ so the kick frequency is larger than the spin Rabi frequency. The spin have not the time to oscillate, to evolve that they are already kicked.

However, for Ising-X or Ising-Z coupling, the evolution with $\frac{w_1}{w_0}$ does not exist here (see fig. \ref{isingzevolutionwithd1} and \ref{driftisingxpopavecd1}). The population and the coherence always go to the microcanonical distribution. The disorder induced by the coupling is so large.

\subsection{Markovian case}
\begin{figure}
\begin{center}
\includegraphics[width=7.7cm]{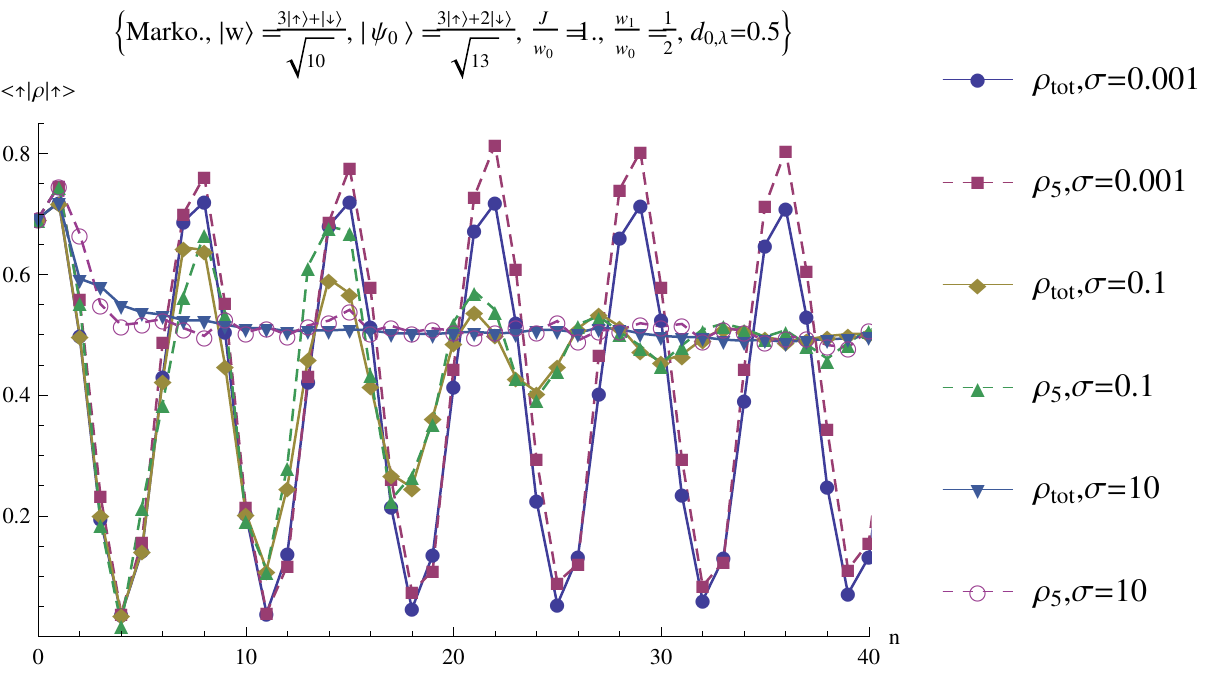}\\
\includegraphics[width=7.7cm]{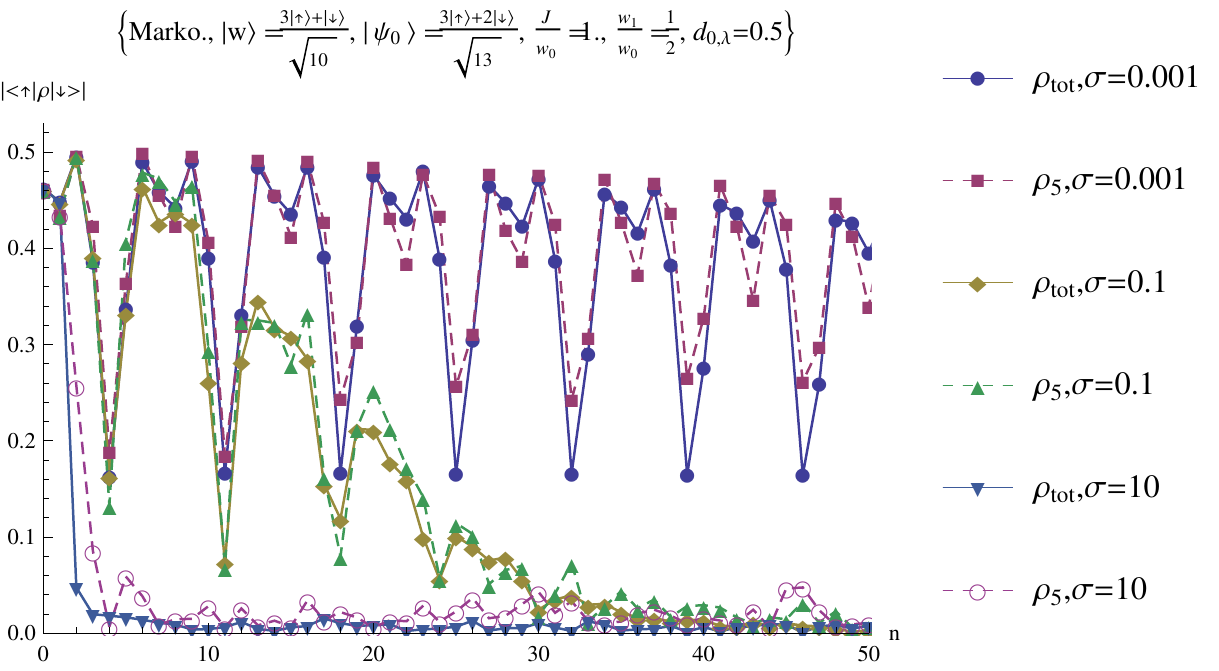}
\caption{\label{gaussheisenberg} Evolutions of the population (up) and of the coherence (down) of a ten spin chain ($\rho_{tot}$) and of the fifth spin of the chain ($\rho_5$) with respect to the standard deviation ($\sigma$) of the Brownian motion. This last defining the evolution of the classical kick bath. The spins are coupled by the Heisenberg interaction.}
\end{center}
\end{figure}

\begin{figure}
\begin{center}
\includegraphics[width=7.7cm]{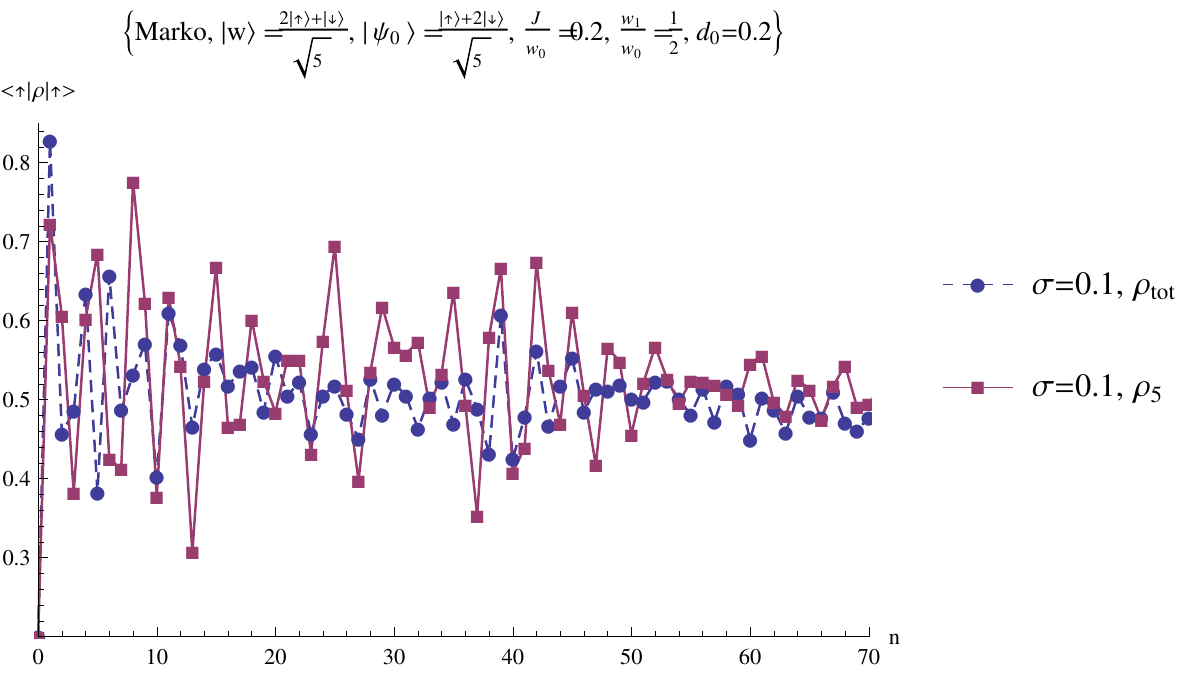}
\caption{\label{isingzentropiemarko} Evolution of the population of a ten spin chain and of the fifth spin of the chain. The spins of the chain are submitted to a Markovian kick bath and are coupled by the Ising-Z interaction. $\sigma$ is the standard deviation.}
\end{center}
\end{figure}

\begin{figure}
\begin{center}
\includegraphics[width=7.7cm]{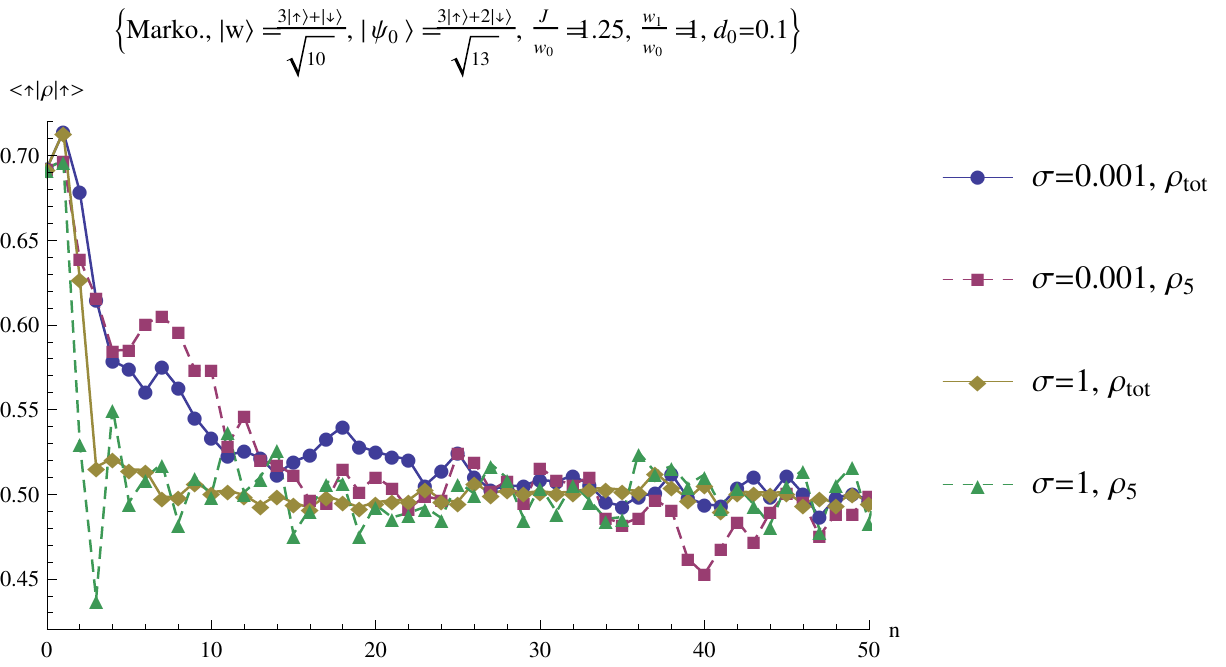}
\caption{\label{markoisingx} Evolutions of the population of a ten spin chain ($\rho_{tot}$) and of the fifth spin of the chain ($\rho_5$). Each spin is submitted to a Markovian kick bath with a small initial dispersion of the kick strengths and of the kick delays ($d_0$). The spins of the chain are coupled by the Ising-X interaction. $\sigma$ is the standard deviation.}
\end{center}
\end{figure}

The Markovian kick bath is special. It depends on the standard deviation $\sigma$. If $\sigma$ is really small, the distribution tends to a stationary one. Unlike, if $\sigma$ is large, the distribution tends to a microcanonical one. Thus, for an average $\sigma$, the distribution begin from a stationary one to a microcanonical one. A damping of the population and coherence oscillation amplitudes is present for a not too large $\sigma$ for an Heisenberg coupling as we can see fig. \ref{gaussheisenberg}.

For Ising-Z, the Markovian kick bath presented fig. \ref{isingzentropiemarko} also shows a population damping (the coherence is also characterized by a damping) if the interaction parameter and the initial dispersion is not too large.

Finally, the entanglement is so large in an Ising-X coupling that it is almost impossible to observe the damping of the population or of the coherence with a Markovian kick bath (fig. \ref{markoisingx}). The fall is very drastic.\\

To sum up, we have just seen that it is possible to not lose all the initial informations because of the disorder, except for the Ising-X coupling. Firstly, we can use a kick in an eigenvector direction for the Heisenberg or the Ising-Z coupling. Secondly, for the Heisenberg interaction, a factor $\frac{w_1}{w_0}$ really small allows to conserve the initial informations. These analyses are interesting for a disordered bath (like a microcanonical one) or when the initial dispersion is large. Finally, for the Heisenberg or for the Ising-Z coupling a Markovian kick bath is interesting to take advantages of the first kick which not induces a decoherence (for a standard deviation not too large). 

Apart from these particular cases, for an ordered kick bath with no large initial dispersion, the Heisenberg coupling does not induce an evolution of the spin behavior to the microcanonical one. Since the spins are kicked similarly, they do not get entangled as we can see on fig. \ref{popcoheheisenbergaveccouplage}.

\section{Conclusion}

\begin{table}
\begin{center}
\caption{\label{baths_behav} Summary of the distinguishing behaviors of a coupled kicked spin chain.}

\begin{tabular}{| p {1.4 cm } | p {2.8 cm }| p {2.8 cm }| p {2.8 cm } |}
\hline
 & \bf Heisenberg & \bf Ising-Z & \bf Ising-X \\
\hline
\hline
$d_0 = 0 $ &  $d_0 <1 \equiv J=0$, no disorder and no entanglement & disorder and entanglement already high if $\frac{J}{w_0}$ is large  & disorder and entanglement already large if $\frac{J}{w_0}$ is large \\
 &  &  & \\
$d_0 \nearrow$ &  disorder and entanglement $\nearrow$ &  disorder and entanglement  $\nearrow$  & disorder and entanglement $\nearrow$ \\
\hline
$\frac{w_1}{w_0} \searrow$ & relaxation $\searrow$ to go to the average weight, the coherence $\nearrow$ with large oscillations & no effect &  no effect \\
\hline
$\frac{J}{w_0} \nearrow $ & disorder and entanglement $\nearrow$  & disorder and entanglement $\nearrow$ & disorder and entanglement $\nearrow$ \\
\hline
$ \vartheta \to 0$ or $\frac{\pi}{2}$ & relaxation $\searrow$ and the spin decoherence & relaxation and decoherence $\searrow$ each spin & nearly no effect \\
 & oscillates more and larger & oscillates more  &  \\
\hline
$\sigma \nearrow $ & decoherence and relaxation $\nearrow$ & decoherence and relaxation $\nearrow$ with small dispersion and interaction &  \\
\hline
particular phenomena & each spin follows the average spin of the chain & a coherence "plateau" & really large entanglement \\
\hline
\hline
\end{tabular}
\end{center}
\end{table}

In this paper, we have studied the behaviors of a coupled spin chain submitted to a kick bath. This bath is disturbed by a classical environment of which the dynamics could be stationary, Markovian, drift or microcanonical. The spins of the chain are coupled by a nearest-neighbor Heisenberg, Ising-Z or Ising-X interaction. The chain behavior with respect to the various system parameters depends on the interaction as summarized Table \ref{baths_behav}. The chain evolution with the Heisenberg interaction is similar of what we have seen without interaction in the paper \cite{viennot2013} for a small initial dispersion. For a large dispersion, the results are similar to ones obtained for a microcanonical kicked bath without interaction. But there is an essential difference : all spins adopt the behavior of the average. Ising-Z interaction induces more disorder than the Heisenberg coupling. But a little initial plateau occurs. The spins try to retain their neighbors before going toward the microcanonical distribution. The interaction which induces the most disorder is the Ising-X coupling. There is always a population relaxation and a decoherence fall toward the microcanonical distribution. 

The disorder into the kick bath, which is due to the classical environment is transmitted to the spin chain. It is larger for an initial dispersion of the kick or a disordered kick bath. If the interaction parameter is still large, it increases the disorder into the spins of the chain through the appearance of the entanglement. So the disorder into the kick bath is transmitted to the chain which is transmitted from one spin to its neighbors through the entanglement. This is really well seen for the Heisenberg coupling with the comparison between a coupled chain and a spin ensemble. A larger spin number in the ensemble is needed to obtain the same general behavior than for a coupled spin chain.

For an ordered kick bath with no large initial dispersion, the Heisenberg coupling does not induce an evolution of the spin behavior to the microcanonical one. Since the spins are kicked similarly, they do not get entangled. For the Heisenberg or for the Ising-Z coupling a Markovian kick bath is interesting to take advantages of the first kick which not induces a decoherence. However for a disordered bath (like a microcanonical one) or when the initial dispersion is large an order conservation can be found. A kick in an eigenvector direction can be used for the Heisenberg or the Ising-Z coupling. For the Heisenberg interaction, a factor $\frac{w_1}{w_0}$ really small allows to conserve the initial informations. These analyses are interesting for a disordered bath (like a microcanonical one) or when the initial dispersion is large.

Now it will be interesting to see the transmission of the information into the coupled spin chain submitted to several kicks. Are there some configurations which allow a transmission of the informations into the spin chain in spite of the disorder of the kick bath and one due to the entanglement between the spins?

\appendix
\section{Discussion about the size of the chain}
\begin{figure}
\begin{center}
\includegraphics[width=7.7cm]{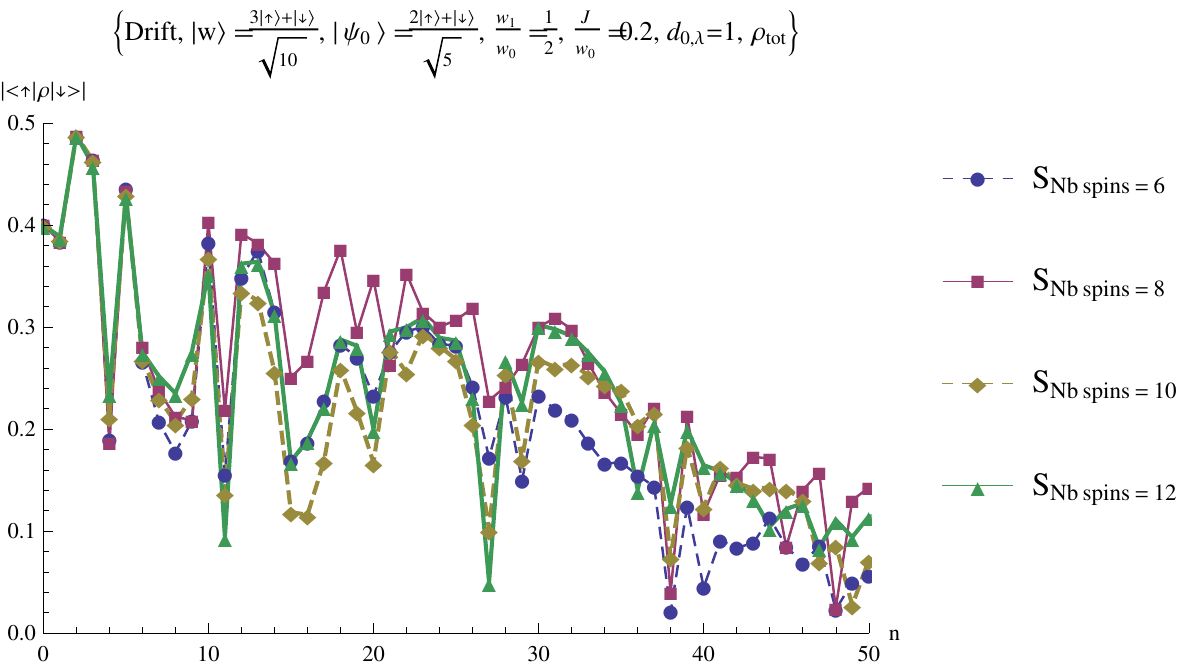}
\includegraphics[width=7.7cm]{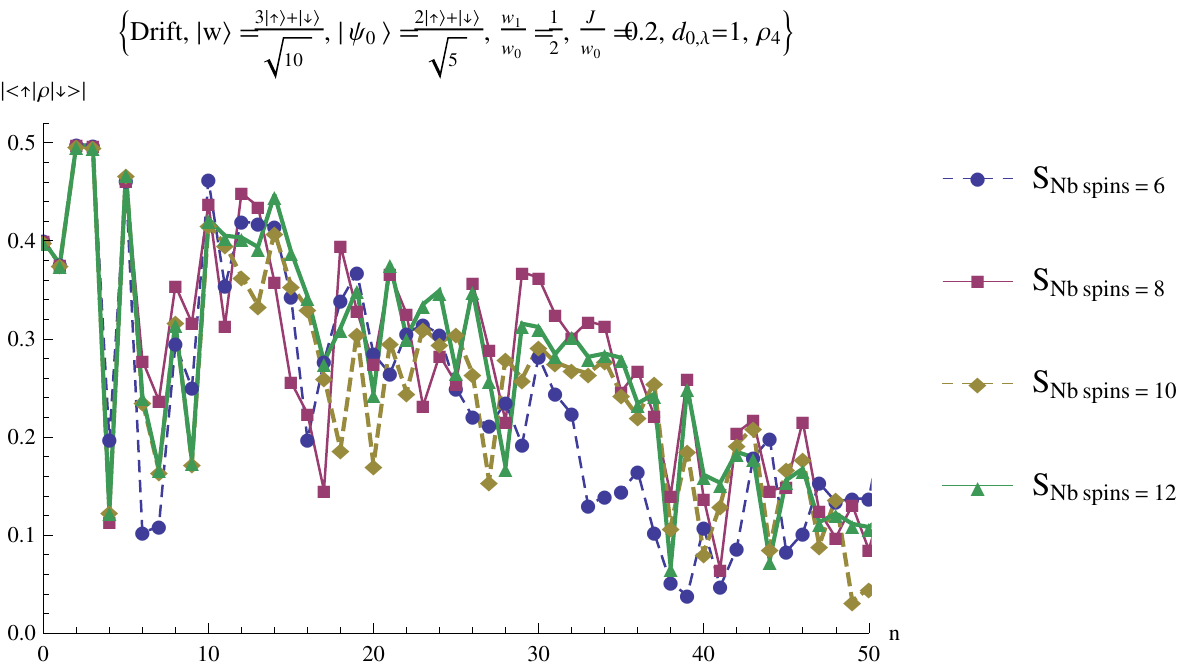}
\caption{\label{heisenbergspinnumber} Evolution of the coherence with various spin numbers into the chain. The spins are coupled to their neighbors by an Heisenberg interaction. Each spin is submitted to a drift classical kick bath. The graphics associated with $\rho_{tot}$ is for the average chain and ones with $\rho_4$ are about the fourth spin of the chain.}

\end{center}
\end{figure} 
\begin{figure}
\begin{center}
\includegraphics[width=7.7cm]{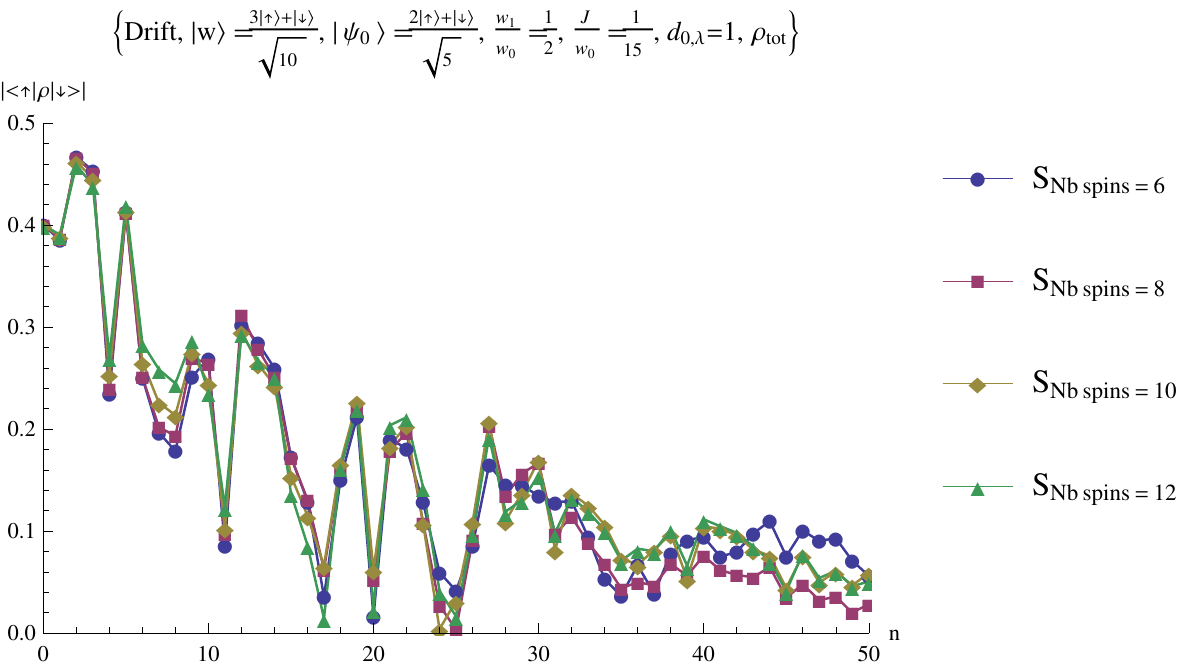}
\includegraphics[width=7.7cm]{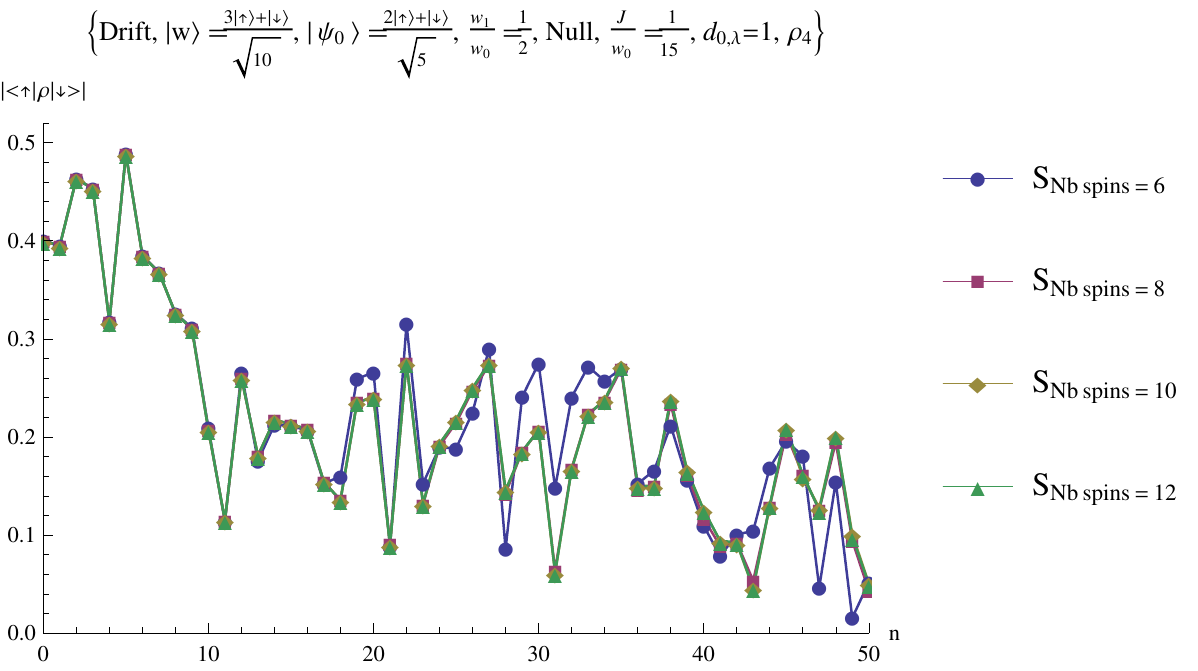}
\caption{\label{isingzspinnumber} Evolution of the coherence with various spin numbers into the chain. The spins are coupled to their neighbors by an Ising-Z interaction. Each spin is submitted to a drift classical kick bath. The graphics associated with $\rho_{tot}$ is for the average chain and ones with $\rho_4$ are about the fourth spin of the chain.}
\end{center}
\end{figure}

\begin{figure}
\begin{center}
\includegraphics[width=7.7cm]{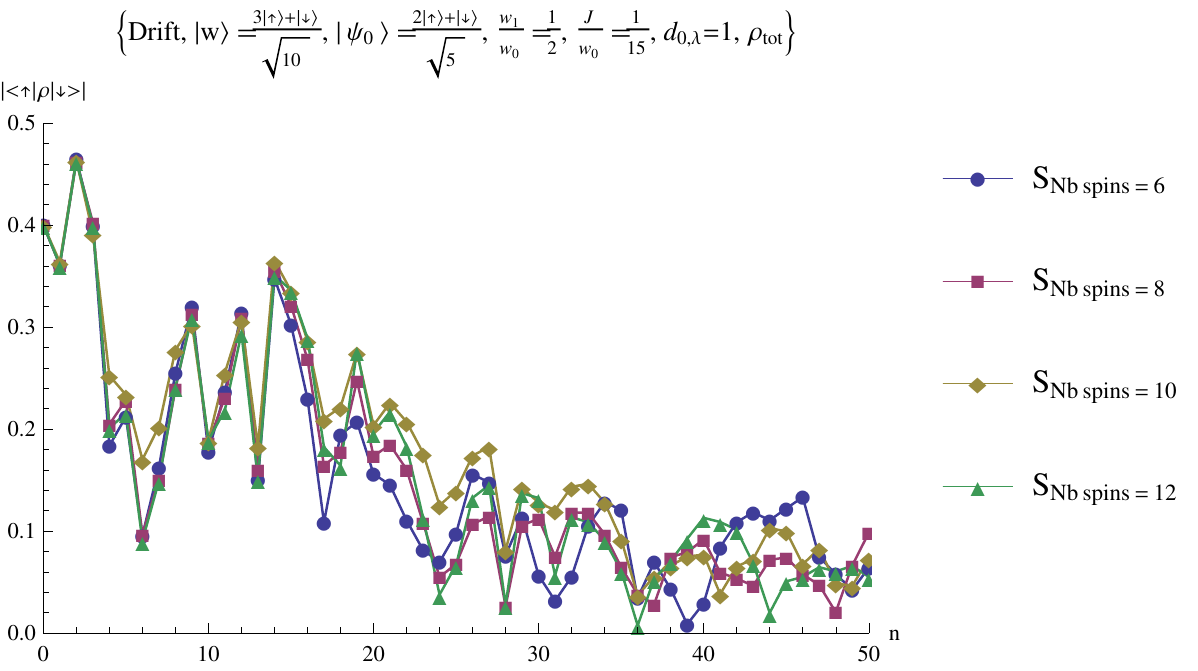}
\includegraphics[width=7.7cm]{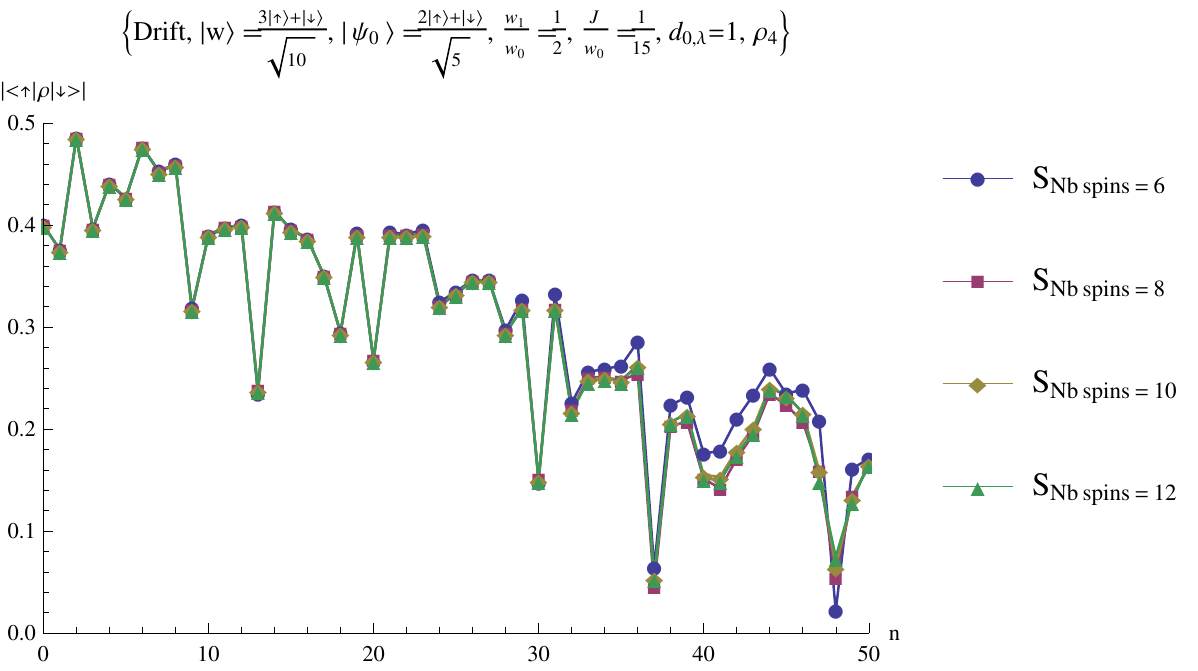}
\caption{\label{isingxspinnumber} Evolution of the coherence with various spin numbers into the chain. The spins are coupled to their neighbors by an Ising-X interaction. Each spin is submitted to a drift classical kick bath. The graphics associated with $\rho_{tot}$ is for the average chain and ones with $\rho_4$ are about the fourth spin of the chain.}
\end{center}
\end{figure}

We have seen the evolution of a ten spin chain coupled by the Heisenberg, Ising-Z or Ising-X interaction and kicked with different initial dispersions. But now, an useful question is to know the influence of the spin number into the chain. To analyze this, 6, 8, 10 and 12 spins have been taken with a medium interaction and with a medium initial dispersion parameters. The coherence is represented for these conditions and for the different couplings fig \ref{heisenbergspinnumber}, \ref{isingzspinnumber} and \ref{isingxspinnumber}. We see on these figures that there is nearly no variation with the spin number.

We can suppose that there is no modification of the spin behavior with the number of spins. So, the analyses will be the same with a large number of spins in the chain.


\begin{thebibliography}{0}
\bibitem{semiclassical} In this analysis, a spin is viewed as a classical magnetic moment vector, inducing a local magnetic field $\vec B_{loc} \propto \langle \vec S \rangle = \tr(\rho \vec S)$ (where $\vec S$ are the spin operators and $\rho$ is spin density matrix) which is felt by their neighbors. In this framework we talk about the (classical) spin orientation in place of the (quantum) spin state (a quantum spin state $\alpha |\uparrow \rangle + \beta |\downarrow \rangle$ being equivalent to the classical spin orientation $\theta = 2 \arctan \left| \frac{\beta}{\alpha} \right|$ and $\varphi = \arg \beta - \arg \alpha$, or in other words we identify the Bloch sphere (the space of the spin normalized states without global phase) characterized by $S^3 = \{ z \in \mathbb{C}^2, ||z|| = 1\}$, with a sphere of classical vector directions).
\end{thebibliography}
\end{document}